\documentclass[
  reprint,
  aps,
  pra,
  amsmath,amssymb,
  floatfix,
]{revtex4-2}

\usepackage{afterpage}
\usepackage{graphicx}
\usepackage{dcolumn}
\usepackage{bm}
\usepackage[tracking=true]{microtype}
\usepackage{braket}
\usepackage[version=4]{mhchem}
\usepackage[dvipsnames]{xcolor}
\usepackage{hyperref}
\usepackage{amsmath, amssymb}
\usepackage[separate-uncertainty=true]{siunitx}
\usepackage{graphicx}
\usepackage{booktabs}
\usepackage{placeins}
\usepackage{tabularx,array} 
\usepackage{titlesec}

\titleformat{\paragraph}[block]
  {\normalfont\normalsize\bfseries\centering} 
  {}                                          
  {0pt}                                       
  {}                                          
\titlespacing*{\paragraph}{0pt}{1.5ex plus .2ex}{0.8ex}

\hypersetup{
	colorlinks=true,       
	linkcolor=blue,  
    citecolor=JungleGreen,        
    filecolor=Orchid,      
    urlcolor=black           
}

\newcommand{\RN}[1]{%
  \textup{\uppercase\expandafter{\romannumeral#1}}%
  }
\usepackage[mathlines]{lineno}
\usepackage{comment}

\begin{document}

\title{Two-photon-excited fluorescence spectroscopy of Rb atoms in a magneto-optical trap} 

\author{Alan McLean$^{1}$}
\author{Christian Drago$^{2}$}
\author{Daniel Podos$^{1}$}
\author{Chengyi Luo$^{1}$}
\author{Caleb Brzezinski$^{1}$}
\author{Ting-Wei Hsu$^{1}$}
\author{J. E. Sipe$^{2}$}
\author{Ralph Jimenez$^{1}$}
\email{Corresponding Author: rjimenez@jila.colorado.edu}

\affiliation{1. JILA, 440 UCB, University of Colorado Boulder, Boulder, CO 80309, USA}

\affiliation{2. University of Toronto, Department of Physics, 60 St George St, Toronto, ON M5S 1A7, Canada}

\begin{abstract}
We report the results of two-photon-excited fluorescence (TPEF) measurements of the $5\mathrm{S}_{1/2} \rightarrow 5\mathrm{D}_{5/2}$ transition of $^{85}$Rb and $^{87}$Rb cooled in a magneto-optical trap (MOT). We observe TPEF at excitation powers as low as 1 $\mu$W or fluxes as low as $2.98_{-0.94}^{+1.37} \times 10^{18}\ \text{photons}\,\text{cm}^{-2}\,\text{s}^{-1}$ ($^{85}$Rb) and $3.31_{-1.33}^{+2.21} \times 10^{18}\ \text{photons}\,\text{cm}^{-2}\,\text{s}^{-1}$ ($^{87}$Rb). Our results demonstrate that optically cooled Rb is a promising platform for observing sensitive two-photon spectral signatures at low photon fluxes.
\end{abstract}

\maketitle
\section{INTRODUCTION \label{sec:introduction}}
Atomic rubidium (Rb) is of interest for optical clocks \cite{Martin2018, Chu2023} and frequency standards \cite{Leng2019, Lu2019}, matter wave interferometers and field sensors \cite{Carey2018, Greve2022, Hu2017, Schmidt2011}, quantum simulation \cite{Chen2023, Luo2025}, and is a resource for  quantum computation \cite{Bluvstein2023}. Recently, Rb has gained attention as a medium suitable for probing the validity of entangled two-photon absorption (ETPA) \cite{Drago2023,Nunez2023,Drago2025}. ETPA occurs when an atom or molecule is driven by correlated photon pairs rather than classical light. The spectral-temporal correlations inherent in the two-photon state of light lead to a linear scaling of a nonlinear process. When compared with classical light, entangled photon pairs may provide a significant enhancement in the low intensity regime. Such an enhancement, if measurable, may be useful for applications relying on two-photon absorption, particularly where photodamage is of concern \cite{Rubart2004, Helmchen2005, bhawalkar1997two}.  

With the success of early experiments by Georgiades et al.  on two-photon absorption driven by narrow-band squeezed light \cite{Georgiades1994} and Dayan et al. on second harmonic generation from a broadband pair source \cite{dayan2005nonlinear}, attention shifted to fluorescent dyes using the broad bandwidth entangled pair sources (either from a bulk or periodically poled crystals) in which the broad bandwidth became a common ``feature" for ETPA. Such studies report detecting ETPA with an enhancement many orders of magnitude above the classical result \cite{tabakaev2021energy, tabakaev2022spatial, villabona2018two, villabona2017entangled, varnavski2017entangled, varnavski2020two, varnavski2022quantum}. These results have been called into question by others who either do not measure an ETPA signal \cite{Parzuchowski2021, landes2024limitations, Raymer:21}, or provide alternative explanations such as hot-band absorption or single-photon scattering for signals which first appear to be ETPA  \cite{Mikhaylov2022, hickam2022single}. Compared with recent experiments using molecular dyes, Rb offers a unique regime of parameter space that can be leveraged to increase the detection capabilities and therefore the measurement of ETPA.   

\begin{figure}[!b]
\includegraphics[width=\linewidth]{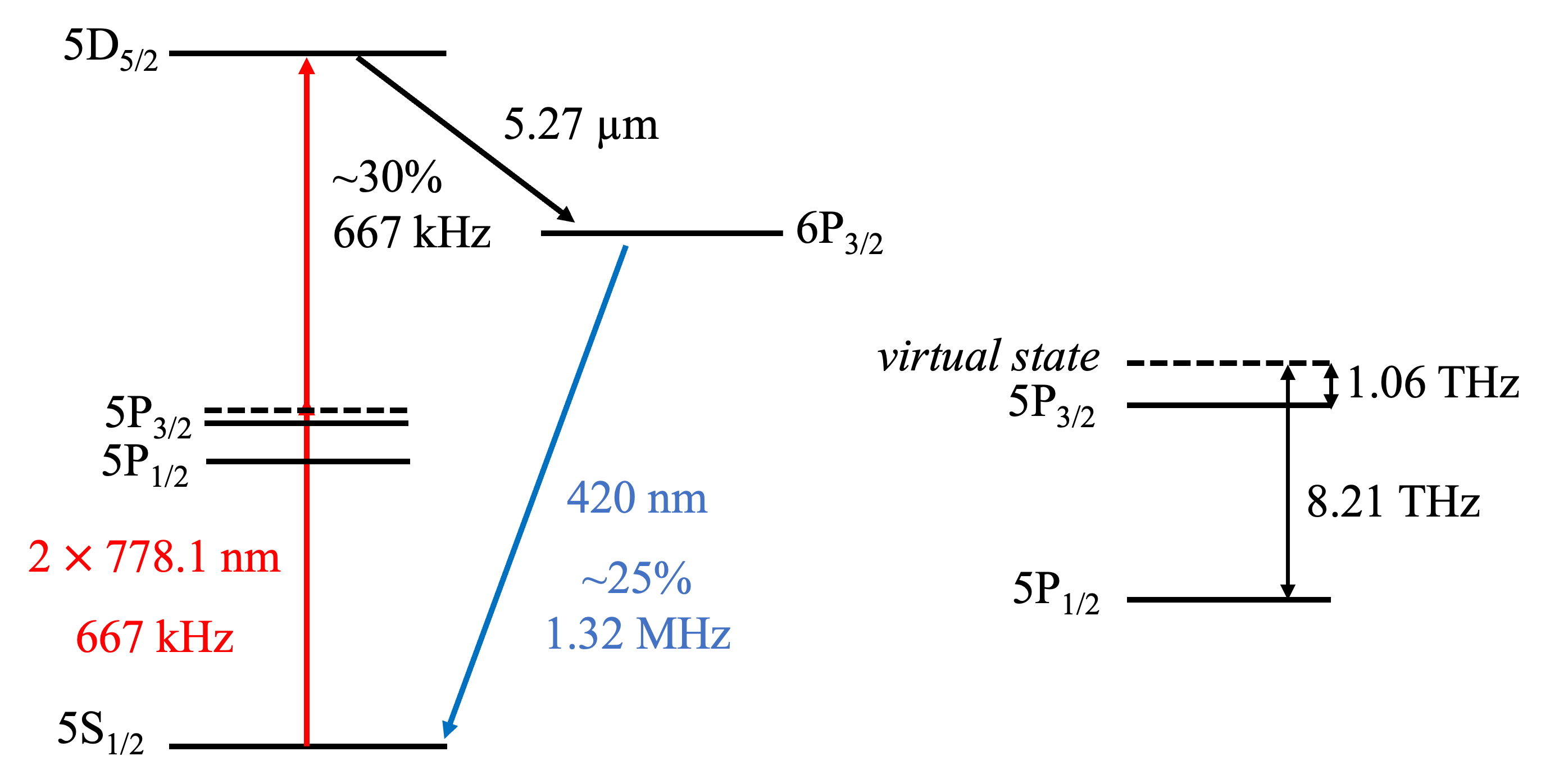}
\caption{The energy level diagram of the $5\mathrm{S}_{1/2} \rightarrow 5\mathrm{D}_{5/2}$ two-photon excitation of Rb (natural linewidth of 667 kHz). Branching ratios and radiative linewidths (calculated from the radiative lifetimes) \cite{Sheng2008} are reported. The narrow two-photon detuning (1.06 THz) between the virtual state and the nearest P state is depicted, in addition to the larger detuning (8.21 THz).}
\label{fig:Jablonski}
\end{figure}

For ETPA, Rb is advantageous because the linewidths of atomic transitions are orders of magnitude smaller than the linewidths of electronic transitions of fluorescent dyes in room temperature solvents. As a result, the two-photon absorption (TPA) cross section is on the order of 10$^{12}$ GM ($1~\mathrm{GM}=10^{-50}$\,cm$^4$\,s\,photon$^{-1}$), which is approximately eight orders of magnitude larger than the cross section for typical molecular systems. This large TPA cross section is of immense relevance for ETPA, since very low intensities are needed to witness an enhancement. Second, when pumped degenerately from the ground state ($5\mathrm{S}_{1/2}$) to the two-photon excited state ($5\mathrm{D}_{5/2}$) the first photon makes a transition near the 1.06 THz detuned real intermediate state, see Fig.~\ref{fig:Jablonski} and Table~\ref{tab:alkali-two-photon} for a comparison across alkali atoms. When compared with continuous wave classical light exciting a degenerate two-photon transition, the broadband nature of the ETPA photons from the SPDC process \cite{Baek2009,Farella2024} leads to spectral overlap and thus resonant excitation of the intermediate state increasing the excitation rate and therefore its quantum advantage  \cite{Drago2023}.

For investigations of ETPA it is crucial that the system under review is well known and can be modeled so that one can accurately predict the results of experiments. For molecular systems this is often difficult as they are typically large complex molecules with many degrees of freedom and where other physical mechanisms such as aggregation may be present. On the other hand, the properties of Rb are extremely well documented due to its many applications in quantum sensing \cite{Carey2018, Greve2022} and precision time and frequency calibrations \cite{Martin2018, Chu2023, Leng2019, Lu2019}. Together these properties make Rb an ideal candidate for making nonlinear quantum-optic calculations.   

Generally there are two types of experiments that are conducted with Rb atoms. Atoms can be in hot vapor cells or can be cooled in a magneto-optical trap (MOT). MOTs allow the confinement of millions of atoms to a small region of space. With a focused excitation beam, this increases the detection capabilities of the experiment for lower incident powers. For example, in an earlier study that motivated our work, Georgiades et al. took advantage of a MOT to study two-photon excitation using a narrow-band nondegenerate squeezed-light source in Cs, showing a linear power dependence to the fluorescent counts \cite{Georgiades1994}.  In his graduate thesis (starting on page 30), Georgiades discussed the advantages of the MOT as compared with the hot vapor cell for two-photon studies, including not requiring the use of counter-propagating beams, MOTs not suffering from Doppler broadening, and the atoms being well-localized \cite{Georgiades1998}. 

Hot vapor systems suffer from Doppler broadening, effectively reducing the rate of excitation by 2-3 orders of magnitude with a classical beam as the bandwidth of light can be far less than the atomic linewidth. For classical TPA, a standard technique to circumvent the limitations due to Doppler broadening is to use counterpropagating beams. On the other hand, photon pairs used for ETPA usually involve correlated degenerate photons with frequencies that add to the same value with a bandwidth much less than the natural linewidth. Consequently, ETPA is uniquely poised to be negatively affected by broadening mechanisms such as Doppler broadening. Since the incident intensities are already very low, these effects may be large enough to suppress any measurable signal \cite{Drago2023}. Further, unlike classical light where photons can be sent into counterpropagating directions, one cannot ``sort'' correlated degenerate photons with equivalent polarizations as in Type-0 or Type-1 crystals such that the photons from a pair are counterpropagating while maintaining the spectral-temporal correlations. Finally, spectral shifts from Doppler broadening which could modulate the absorption rate \cite{Georgiades1998} and ``incoherent'' two-photon absorption of the photon pairs \cite{Drago2023} must be considered in hot vapors. In a MOT, however, the incoherent TPA rate is negligible due to the absence of substantive Doppler broadening \cite{Drago2023}.

{To improve detection capabilities and remove the effects of Doppler broadening we present an analysis of Rb driven by classical TPA in a MOT cooled down to temperatures in the $\mu$K regime.} Despite its potential, there are few reports of two-photon-excited fluorescence (TPEF) from ultracold Rb atoms \cite{Leng2019, Lu2019}. Previous works have reported TPEF of ultracold Rb down to 10 mW powers, which correspond to fluxes on the order of 10$^{20}$ cm$^{-2}$ s$^{-1}$ \cite{Leng2019, Lu2019}. In this report, we demonstrate TPEF power-dependence of ultracold Rb down to excitation powers of $\sim$1 $\mu$W or fluxes as low as 2.98 $\times$ 10$^{18}$ cm$^{-2}$ s$^{-1}$. This sensitivity in flux is better than any previously published atomic two-photon study by over a factor of 10-fold, or a factor of $>$100-fold in terms of counts, since the TPEF counts scale quadratically with power. Additionally, this sensitivity is approximately $5\times$ greater than that reported in the best molecular ETPA studies to date \cite{Mikhaylov2022}. Our results demonstrate that the Rb MOT is a promising platform for low-flux TPEF to be used for observing novel non-linear phenomena like ETPA.

The outline of this paper is as follows: In Sec. \ref{sec:methods} we provide an outline of our procedure with results provided in Sec. \ref{sec:results and discussion}. Our concluding remarks are presented in Sec. \ref{sec:Conclusion} with various supplementary information provided in appendices. 

\begin{figure}[!b]
\includegraphics[width=\linewidth]{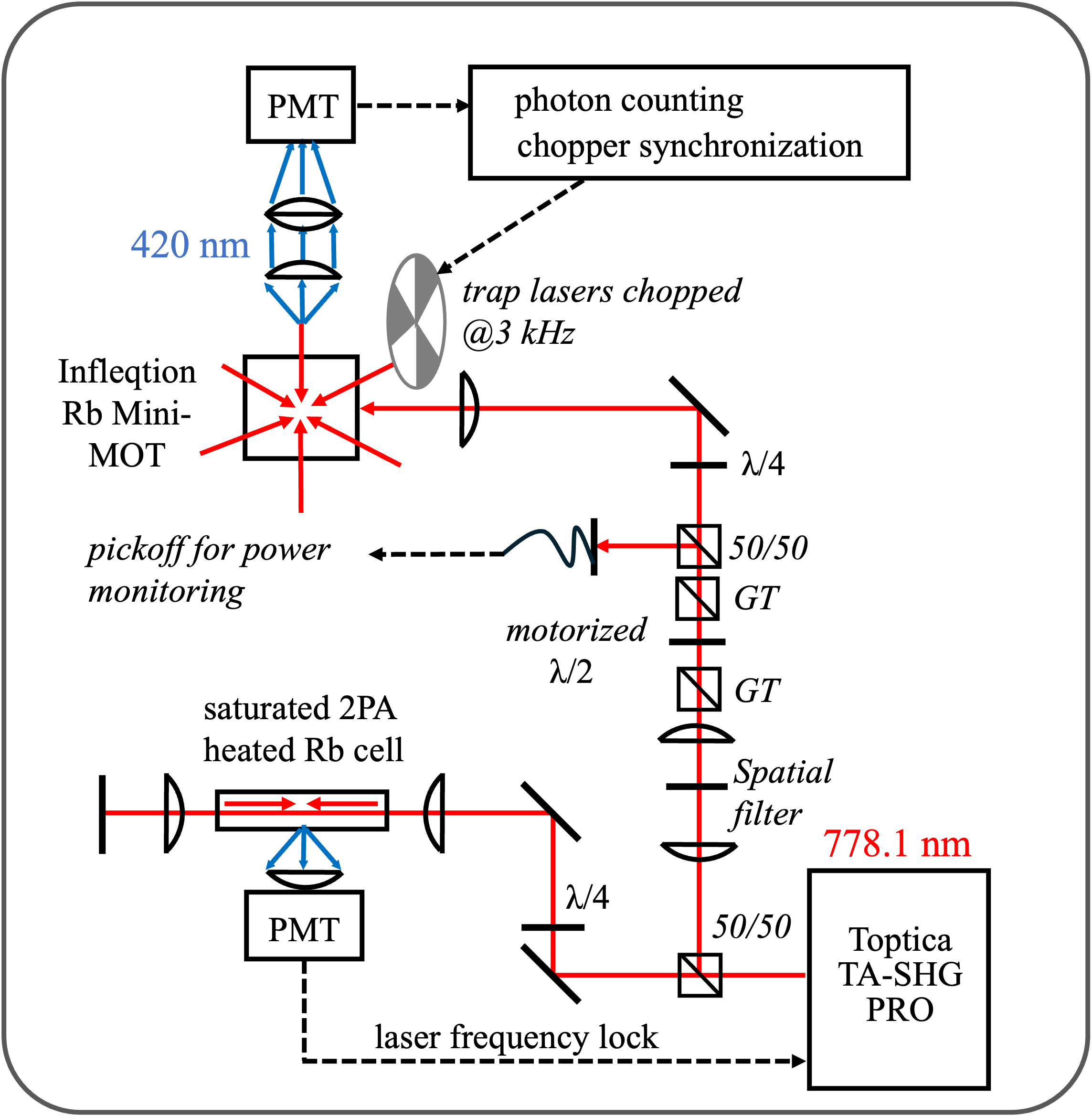}
\caption{Schematic for two-photon excitation of the Rb MOT. GT = Glan-Taylor polarizer.}
\label{fig:two-photon-schematic}
\end{figure}

\section{Experimental Overview \label{sec:methods}}
Here we provide an overview of our experiment (schematic shown in Figure~\ref{fig:two-photon-schematic}). An Infleqtion mini-MOT V2 dispenses Rb (natural isotope abundance) at an ultralow pressure (nTorr) into a sealed, optically transparent glass cell. Taking advantage of the natural isotope composition, we created MOTs of both $^{85}$Rb and $^{87}$Rb. In brief, trap and repump diode lasers with frequencies either locked to $^{85}$Rb and $^{87}$Rb (see details in Appendix~\hyperref[app:details-MOT]{A}) were sent into the MOT to optically cool and keep atoms in a single hyperfine ground state ($5S_{1/2}\,F_g = 3$ for $^{85}\mathrm{Rb}$, and $5S_{1/2}\,F_g = 2$ for $^{87}\mathrm{Rb}$). We used an optical chopper to activate and deactivate the trap and repump beams with a 50-50 duty cycle, providing a time window for probing the MOT when all atoms were in the ground state (when the beams were disengaged). This cycle produced a steady-state MOT. The number of trapped atoms in the MOT was determined via absorption using a separate probe laser for imaging during the time window when the trap and repump beams were blocked  (see Appendix~\hyperref[app:details-absorption-imaging]{A}). 

Two-photon excitation of the MOT was performed using a laser (778.1 nm) locked via Doppler-free two-photon spectroscopy to the strongest two-photon transition of each isotope $^{85}\mathrm{Rb}\,(5\mathrm{S}_{1/2},\,F_g = 3 \rightarrow 5\mathrm{D}_{5/2},\,F_f = 5)$ and $^{87}\mathrm{Rb}\,(5\mathrm{S}_{1/2},\,F_g = 2 \rightarrow 5\mathrm{D}_{5/2},\,F_f = 4)$. Using a motorized waveplate controller and Glan-Taylor polarizer, the power of the beam could be varied for power dependence studies. After the Glan-Taylor polarizer, a 50-50 beam splitter was used as a pickoff to monitor the beam power, while the remainder of the beam was sent into the MOT. After the beam splitter, a quarter-waveplate was used to create circularly polarized light. Fluorescence was collected orthogonally to the excitation using a custom-designed 2-lens system that focused light into a photon-counting PMT. The fluorescence signal was time-gated (synchronized with the optical chopper) to analyze data only when the trap and repump beams were blocked. The TPEF experiment was fully automated using 5 s on (excitation beam delivered)-off (excitation beam blocked) cycles. A block averaging time of 400 s of the two-photon fluorescent counts was selected by taking Allan deviation measurements (Appendix~\ref{app:allan-deviation}, Figs.~\ref{fig:allan-mot} and \ref{fig:allan-dark}). Additionally, the total collection time was capped at 4 hours as the Allan deviation of the PMT dark counts (evaluated with the lasers used to make the MOT) deviated from linearity beyond that time. Data processing was performed using hand-coded Python scripts for all measurements (two-photon power dependence studies, absorption imaging, etc.).

\section{Results and discussion \label{sec:results and discussion}}

\begin{figure*}[!htbp]
  \centering
  \includegraphics[width=\linewidth]{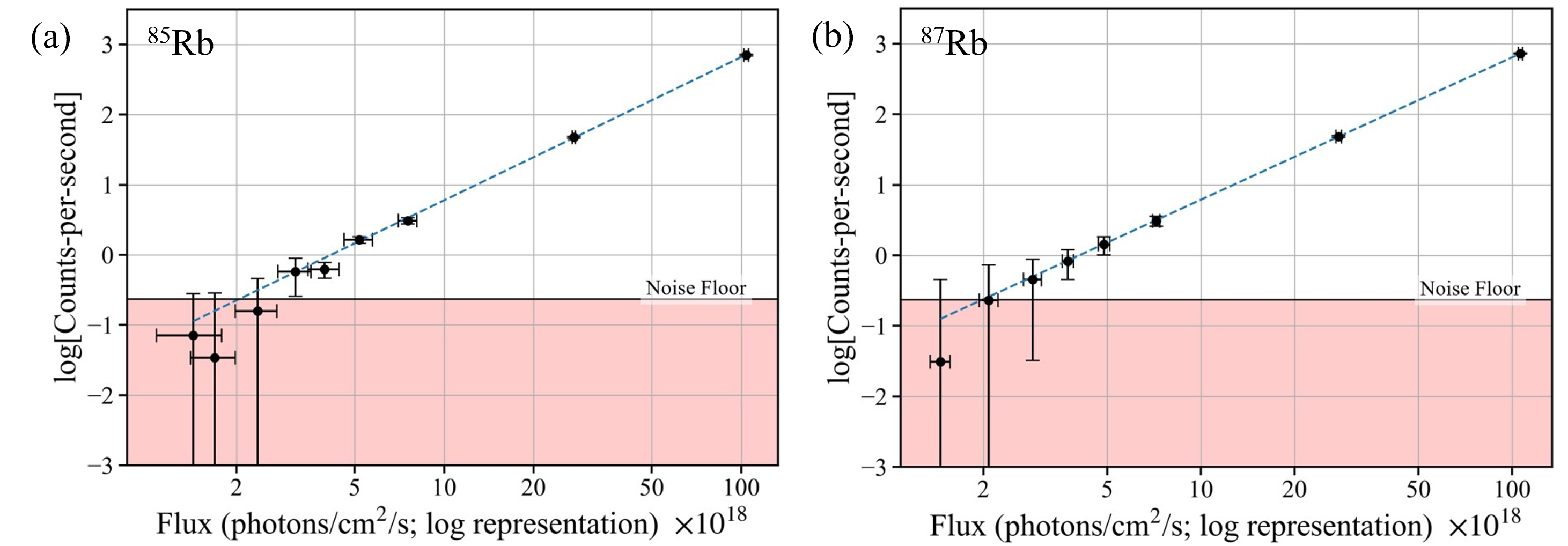}
  \caption{Log--log power-dependence plots for (a) $^{85}$Rb and (b) $^{87}$Rb. Linear fits of the form $y = m x + b$ in $\log_{10}$--$\log_{10}$ space give, for $^{85}$Rb: $m = 2.041 \pm 0.023$, $b = -38.000 \pm 0.445$; and for $^{87}$Rb: $m = 2.018 \pm 0.004$, $b = -37.554 \pm 0.078$. A slope consistent with $m \approx 2$ indicates fluorescence arising purely from two-photon processes. The noise floor is indicated by the solid black line; points whose error bars cross this line are not statistically significant.}
\label{fig:log-log}
\end{figure*}

Fluorescence collection methods are less prone to error, and can achieve higher sensitivities, than absorption approaches (as demonstrated by our group, up to a factor 10$^{3}$ to 10$^{4}$ better in the context of the two-photon cross section) \cite{Parzuchowski2023}. We thus utilized the two-photon excited fluorescence (TPEF) method \cite{Xu1996} to determine the two-photon absorption cross section of Rb. The TPEF method involves measuring the power-dependent fluorescence of the sample \cite{Xu1996}. Power-dependent TPEF spectroscopy was performed on the trapped Rb atoms. For both $^{85}$Rb and $^{87}$Rb, the log-log plot of the power dependence had a slope of 2.0, within statistical significance, Figure~\ref{fig:log-log}. This indicates that the fluorescence originates purely from TPA. 

Using the y-intercept values from the fit, and parameters from the experiment (such as the atom number, MOT dimensions, and beam waist; full parameters shown in Appendix~\ref{app:two-photon-params}), the two-photon cross section ($\delta$) of each Rb isotope was calculated. For $^{85}$Rb, the two-photon cross section was $1.07\pm0.20\times10^{12}$ GM, Table~\ref{tab:tp-cross sections}. The $^{87}$Rb two-photon absorption cross section was $2.64\pm0.49\times10^{12}$ GM, Table~\ref{tab:tp-cross sections}.

\begin{table}[b!]
\centering
\caption{Experimental two-photon cross sections in GM ($1~\mathrm{GM}=10^{-50}$\,cm$^4$\,s\,photon$^{-1}$) for Rb 5S$\!\to\!$5D$_{5/2}$.
All reports in Table~\ref{tab:tp-cross sections} used the two-photon excited fluorescence to determine the cross section. Saha et al. and Zapka et al. used linearly polarized excitation in their studies. In this study, we used circularly polarized excitation, which increases the two-photon excitation rate up to $1.75\times$. Detailed measurement parameters for our work are given in Table~\ref{tab:two-photon-params}.}
\label{tab:tp-cross sections}
\renewcommand{\arraystretch}{2}
\newcolumntype{Y}{>{\raggedright\arraybackslash}X}
\begin{tabularx}{\linewidth}{Y c Y Y}
    \hline\hline
    \textbf{Reference} & \textbf{Isotope} & \textbf{Method} & \textbf{$\delta$ (GM)}\\
    \hline
    Saha et al. \cite{Saha2011} & $^{85}$Rb & Fiber \newline linear pol. & $1.28~\times~10^{12}$\\
    Zapka et al. \cite{Zapka1983} & $^{87}$Rb & Hot vapor \newline linear pol. & $1.0~\times~10^{12}$\\
    \textit{This work} & $^{85}$Rb & MOT \newline circular pol. & \textit{measured} $(1.07~\pm~0.20)~\times~10^{12}$\\
    & & & \textit{natural} $(2.40~\pm~0.49)~\times~10^{12}$\\
    \textit{This work} & $^{87}$Rb & MOT \newline circular pol. & \textit{measured} $(2.64~\pm~0.49)~\times~10^{12}$\\
    & & & \textit{natural} $(5.94~\pm~1.17)~\times~10^{12}$\\
    \hline\hline
\end{tabularx}
\end{table}

The two-photon cross section measurement has four main sources of error: the log-log fit intercept, the spot size, the atom density, and the collection efficiency, Appendix~\ref{app:two-photon-params}. The statistical uncertainty of the log--log fit is quantified by the
$1\sigma$ uncertainty of the fitted $y$-intercept, yielding $6\%$ for $^{85}\mathrm{Rb}$ and $1\%$ for $^{87}\mathrm{Rb}$ (for this calculation the plus-minus value is first exponentiated to linear space, then an average of the percent error is reported). Fitting in log-space amplifies uncertainties, and this error is small relative to the other parameters. For example, the focused spot size, calculated from the beam dimensions before the focus and taking into account the M$^2$, has 10$\%$ error. The atom density also has $\sim$10$\%$ error, which largely stems from error in determining the atom number (calculated via absorption imaging). Given the OD range of our experiment (up to ODs of 5), 10$\%$ error is expected \cite{Hoekstra2017} and would require advanced techniques to reduce further \cite{Reinaudi2007, Hueck2017}. Finally, the collection efficiency error (12.4$\%$) is determined from Zemax modeling, Appendix~\ref{app:fluorescence-efficiency}. We designed the fluorescence collection with 2 lenses to operate over wide offsets with minimal effect on the collection efficiency (Appendix \ref{app:fluorescence-efficiency}), and the error is comparable to other MOT experiments \cite{Raab1987}.

In addition to error sources, since the measured two-photon cross section depends on the on-resonance fluorescence rate, it is significant that the MOT two-photon excitation linewidth should be considered in interpreting the cross section values. The observed two-photon linewidths in the MOT ($1.5\pm0.1$~MHz, Appendix~\ref{app:MOT-excitation-spectra}) exceed its natural 5D$_{5/2}$ linewidth of $667$~kHz. This is likely due to magnetic-field effects that arise when the MOT coils remain on during the probe, of which we provide calculations in Appendix~\ref{app:broadening-calcs}. In this configuration the atoms occupy a distribution of magnetic sublevels $m_F$, each of which experiences a different Zeeman shift in the quadrupole magnetic field. Even within the small excitation volume set by the tightly focused 778~nm beam ($w_0 \approx 6.5~\mu\mathrm{m}$, $z_R \approx 0.17~\mathrm{mm}$), the applied gradient of $\sim 14~\mathrm{G/cm}$ produces a field variation of order $0.2$~G across the interaction region. When combined with the finite local field ($\sim 0.1$--$0.2$~G), this leads to a corresponding multi-component Zeeman manifold in which different $m_F \rightarrow m_{F''}$ two-photon pathways are shifted by several hundred kilohertz relative to one another. Smaller contributions from residual Doppler broadening, pressure broadening, and power broadening also add in quadrature, but these effects are minor compared with the magnetic field-induced broadening, Appendix~\ref{app:broadening-calcs}.

Although the broadening would be dominated by inhomogeneous mechanisms, it is useful to estimate an effective TPA cross section that would be inferred if the full linewidth were treated as homogeneous broadening of a single Lorentzian. Under that assumption the on-resonance response scales inversely with the linewidth, so rescaling from $\Gamma_{\mathrm{MOT}} = 1.5\pm0.1$~MHz (Appendix~\ref{app:MOT-excitation-spectra}) to the natural 5D width $\Gamma_{\mathrm{nat}} = 0.667$~MHz \cite{Sheng2008} multiplies the extracted cross section by the factor $\Gamma_{\mathrm{MOT}}/\Gamma_{\mathrm{nat}}~= 2.3\pm0.2$. Applying this factor with error propagation to our measured values yields natural-width upper limits $\delta_{85}^{(\mathrm{nat})} \approx 2.40 \pm 0.49\times 10^{12}~\mathrm{GM}$ and $\delta_{87}^{(\mathrm{nat})} \approx 5.94 \pm 1.17\times 10^{12}~\mathrm{GM}$, where these values should be interpreted as upper bounds, Table~\ref{tab:tp-cross sections}.

To our knowledge, only two prior experimental studies have reported two-photon cross sections for the same 5S$\rightarrow$5D transitions in rubidium, both using linearly polarized excitation, Table~\ref{tab:tp-cross sections} \cite{Zapka1983, Saha2011}. Saha et al.\ report a value of $\delta = 1.28\times 10^{12}$~GM for $^{85}$Rb in a fiber-guided geometry, while Zapka et al.\ find $\delta \simeq 1.0\times 10^{12}$~GM for $^{87}$Rb in a vapor cell. Both earlier works employ analysis frameworks that normalize the extracted cross section to the natural homogeneous linewidth (including transit-time and Doppler contributions in the fiber experiment). The natural linewidth two-photon cross section estimates lie in the same $10^{12}$~GM range as these earlier $^{85}$Rb and $^{87}$Rb studies, Table~\ref{tab:tp-cross sections}. Finally, we used circularly polarized excitation, which increases the two-photon excitation rate 1.75-fold \cite{Olson2006}, and would contribute to the higher cross section bounds. 

\begin{table*}[htpb!]
\centering
\caption{Minimum detectable power $P_{\min}$ and corresponding photon flux $\Phi_{\min}$ for atomic two-photon excitation experiments in Rb and Cs.}
\label{tab:tp-flux}
\renewcommand{\arraystretch}{2}
\begin{tabularx}{\linewidth}{Xc X X X X}
    \hline\hline
    \textbf{Reference} & \textbf{Species} & \textbf{Platform} & \textbf{Type} & \textbf{P$_{min}$} & \textbf{$\phi$$_{min}$} \\ 
    &&&&&(photons cm$^{-2}$ s$^{-1})$\\
    \hline
    Caracas N\'{u}\~{n}ez et al. \cite{Nunez2023} & $^{133}$Cs & Hot vapor fluorescence & power dependence & 10 mW & $1.63 \pm 0.11 \times 10^{22}$ \\
    Saha et al. \cite{Saha2011} & $^{85}$Rb & Fiber fluorescence & power dependence & 200 $\mu$W & $7.8 \times 10^{21}$\\
    Zapka et al. \cite{Zapka1983} & $^{87}$Rb & Hot vapor fluorescence &  operating & 60 mW & $1.6 \times 10^{21}$\\
    Collins et al. \cite{Collins1993} & $^{85}$Rb & Hot vapor absorption & operating & 4.7 mW & $6.8 \pm 0.2 \times 10^{19}$\\
    Hassanin et al. \cite{Hassanin2023} & $^{87}$Rb & Hot vapor IR emission & operating & 25 mW & $5.5 \times 10^{20}$\\
    Leng et al. \cite{Leng2019} & $^{87}$Rb & MOT fluorescence & operating & 20 mW & not applicable\\
    \textit{This work} & $^{85}$Rb & MOT fluorescence & power dependence & $1.01_{-0.32}^{+0.46}~\mu\mathrm{W}$ & $2.98_{-0.94}^{+1.37} \times 10^{18}$\\
    \textit{This work} & $^{87}$Rb & MOT fluorescence & power dependence & $1.12_{-0.45}^{+0.74}~\mu\mathrm{W}$ & $3.31_{-1.33}^{+2.21} \times 10^{18}$\\
    \hline\hline
\end{tabularx}
\begin{flushleft}
\textit{Notes:} Saha \emph{et al.} used two counter-propagating 100\,\textmu W beams (total = 200\,\textmu W) in a hollow-core fiber \cite{Saha2011}. Collins \emph{et al.} used \textbf{$P_{+}=2.9\pm0.1$}\,mW and
\textbf{$P_{-}=1.8\pm0.1$}\,mW for each beam \cite{Collins1993}. Lu et al. (2019) \cite{Lu2019} (not included in table) published a study with the same parameters as Leng et al. \cite{Leng2019}.
\end{flushleft}
\end{table*}

Having measured the two-photon cross section of the cold Rb atoms, the sensitivity (the lowest power and fluxes to produce a signal with confidence taking error into account) was determined. The power sensitivity of the measurement above the noise floor was $1.01_{-0.32}^{+0.46}$ $\mu$W ($^{85}$Rb) and $1.12_{-0.45}^{+0.74}$ $\mu$W ($^{87}$Rb) respectively, Table~\ref{tab:tp-flux} and Appendix~\ref{app:log-log-power}, Fig.~\ref{fig:power-dependence}. These limits are $>$100x lower than was demonstrated in fiber \cite{Saha2011}, more than 1000x lower than has been demonstrated in alkali hot vapor \cite{Nunez2023}, and more than 1000x lower than previously demonstrated in an ultracold MOT \cite{Leng2019}, Table~\ref{tab:tp-flux}. 

For $^{85}$Rb and $^{87}$Rb, the flux sensitivity was $2.98_{-0.94}^{+1.37} \times 10^{18}\ \text{photons}\,\text{cm}^{-2}\,\text{s}^{-1}$ and $3.31_{-1.33}^{+2.21} \times 10^{18}\ \text{photons}\,\text{cm}^{-2}\,\text{s}^{-1}$ respectively. When compared between previous Rb studies (Table~\ref{tab:tp-flux}) and across ETPA studies \cite{Parzuchowski2025, Parzuchowski2021, Mikhaylov2022}, the sensitivity in flux is better than other previously published studies (for ETPA studies nearly 5x improved) \cite{Mikhaylov2022}. While Rb density in the MOT ($2.22\times10^{10}~\mathrm{cm}^{-3}$) is substantially lower than molecular systems (for example, a 1 mM molecular concentration corresponds to $6.02\times10^{17}~\mathrm{cm^{-3}}$), the $>$10$^{10}$ larger cross section pushes the measurement sensitivity into a new regime.

Our results indicate that ultracold Rb is a promising model platform for studying low-flux two-photon effects like ETPA. For example, using periodically poled KTP (ppKTP), we have produced up to 750 nW of Type-0 SPDC when using a 389 nm seed (doubled from the 778.1 nm frequency-locked laser). This result, consistent with reports from literature, would produce a flux of $2.22\times10^{18}$ photons cm$^{-2}$ s$^{-1}$ assuming an equivalent spot size to the classical beam. Using two SPDC crystals or double-passing SPDC photons through the MOT would push the flux above our limit of detection. In practice, however, focusing of a highly polychromatic source like SPDC is challenging (for example, at present, we have found we can focus the SPDC to $\sim$50 $\mu$m). Future work is aimed to systematically study and optimize SPDC, and in tandem with theory, to measure an ETPA signal in the MOT platform.  

\section{CONCLUSION \label{sec:Conclusion}}

In conclusion, we studied two-photon excitation of Rb in a magneto-optical trap (MOT). Log-log power dependence studies confirmed fluorescence purely from two-photon absorption in the Rb MOT as indicated by the fitted slope of 2. We achieved $2.98_{-0.94}^{+1.37} \times 10^{18}\ \text{photons}\,\text{cm}^{-2}\,\text{s}^{-1}$ ($^{85}$Rb) and $3.31_{-1.33}^{+2.21} \times 10^{18}\ \text{photons}\,\text{cm}^{-2}\,\text{s}^{-1}$ ($^{87}$Rb) minimum flux sensitivity. With a flux sensitivity over an order of magnitude better than other published systems (and $\sim$5x improved over deliberately optimized molecular ETPA studies), we find that ultracold Rb is a very promising platform for observing low-flux two-photon signatures. 

\begin{acknowledgments}
This work was supported by the NSF Physics Frontier Center at JILA (PHY 2317149) and the Natural Sciences and Engineering Research Council of Canada (NSERC). AM acknowledges support from the NRC research associateship programs (RAP), 2023-2025. We thank James Thompson for his advice in planning these experiments. We thank James Thompson and Adam Kaufman for feedback on the manuscript. AM thanks JILA electronics staff Terry Brown, Ivan Ryger, and James Fung-A-Fat, and all JILA machine shop staff. AM thanks Dr. Kristen Parzuchowski for her initiation and guidance at the start of his time at JILA, as well as for valuable discussions. He also thanks Miles San Soucie and Ryan Benson for their help in the laboratory and for valuable discussions. AM also thanks Noah Lindsell and Dr. Paul Kunz for their feedback and discussion on the arXiv preprint, which improved the quality of the final manuscript.  

Certain commercial equipment, instruments, or materials are identified in this paper in order to specify the experimental procedure adequately. Such identification is not intended to imply recommendation or endorsement by NIST, nor is it intended to imply that the materials or equipment identified are necessarily the best available for the purpose.
\end{acknowledgments}

\section*{DATA AVAILABILITY}
The data that support the findings of this article are not publicly available. The data are available from the authors upon reasonable request.

\appendix
\titleformat{\section}[block]
  {\normalfont\bfseries\centering}  
  {APPENDIX~\Alph{section}.}        
  {0.5em}                           
  {}                                

\section{DETAILS OF EXPERIMENT}
\subsubsection{Cold Rb magneto-optical trap (MOT)}
\label{app:details-MOT}
Using a standard six-beam geometry \cite{Rushton2014}, we side-locked two near-infrared (NIR) lasers to the hyperfine transitions for cooling (5$\mathrm{S}_{1/2}$ F = 3 $\to 5\mathrm{P}_{3/2}$ $\mathrm{F}^\prime$ = 4) and re-pumping (5$\mathrm{S}_{1/2}$ F = 2 $\to$ 5P$_{3/2}$ $\mathrm{F}^\prime$ = 3) of $^{85}$Rb, respectively. For $^{87}$Rb, the corresponding cooling and re-pumping transitions are 5S$_{1/2}$ F = 2 $\to$ 5P$_{3/2}$ $\mathrm{F}^\prime$ = 3 and 5S$_{1/2}$ F = 1 $\to$ 5P$_{3/2}$ $\mathrm{F}^\prime$ = 2. The locking of both lasers was achieved using PII (`proportional-integral-integral') loop filters, and the MOT isotope (85 and 87) was readily selectable by tuning our lasers to the proper transitions.

In order to selectively excite ground-state Rb atoms (as when actively cooled the Rb atoms are optically driven between the 5S$_{1/2}$ and 5P$_{3/2}$ states), we employed an optical chopper set at 3 kHz to rapidly disengage and re-engage the cooling and re-pump lasers. With the optical chopper engaged, we trapped $>$10$^{7}$ atoms in a steady-state, which we quantified via absorption imaging of the MOT. By synchronizing the optical chopper with time-binned fluorescence detection (Hamamatsu H10682-210 PMT, PicoQuant HydraHarp 400), we probed the TPEF of the ultracold Rb atoms when all atoms were in the ground state, i.e. when the optical chopper was blocking the cooling and re-pump MOT beams. 

\subsubsection{Two-photon Excitation Source (Locked Standard)}
\label{app:details-two-photon-excitation}
\indent In the ultracold MOT, Doppler-broadening is negligible, allowing for Doppler-free excitation of Rb. Frequency stabilization is thereby required in order to ensure that the laser used for two-photon excitation is on-resonance. Feedback control (peak locking) was used to maintain resonant laser excitation. To create the two-photon lock, we used a Toptica TA-SHG pro laser (778.1 nm emission) arranged in a Doppler-free configuration.  We collected the TPEF using a two lens-filter-PMT design orthogonal to the hot Rb cell. The Rb cell was heated to $\approx$ $90^\circ$ C. 

Heating the Rb vapor cell increases the atomic density in the interaction region \cite{Pizzey2022}. Temperatures of $\approx 90^{\circ}\mathrm{C}$ are typically required to obtain a measurable signal \cite{Olson2006,Terra2016}, whereas heating beyond $\sim 125^{\circ}\mathrm{C}$ provides no further improvement in borosilicate cells and may gradually deplete the vapor through alkali--glass bonding \cite{Ryan1993,Olson2006}. We therefore operated the cell only as hot as necessary for stable signal generation.

The cell was enclosed in a copper sheath with a side window for fluorescence collection. A $60$~W silicone resistive heater was wrapped around the sheath and secured with worm--gear clamps. To prevent Rb condensation on the entrance and exit faces, custom copper end caps provided thermal contact to both windows, each containing a $2$~mm beam aperture.

The assembly was supported by an Ultem cradle mounted on a three--axis translation stage. Ultem, rated above $170^{\circ}\mathrm{C}$ \cite{SABIC_Ultem1000}, provides both mechanical rigidity and thermal isolation. Insulation performance was verified using $30$~k$\Omega$ thermistors placed on the cell wall and within the cradle. After four hours at $50$~W heater power, the cell stabilized at $140^{\circ}\mathrm{C}$ while the cradle remained within $1^{\circ}\mathrm{C}$ of ambient, confirming effective isolation.

\begin{figure}[b!]
\includegraphics[width=\linewidth]{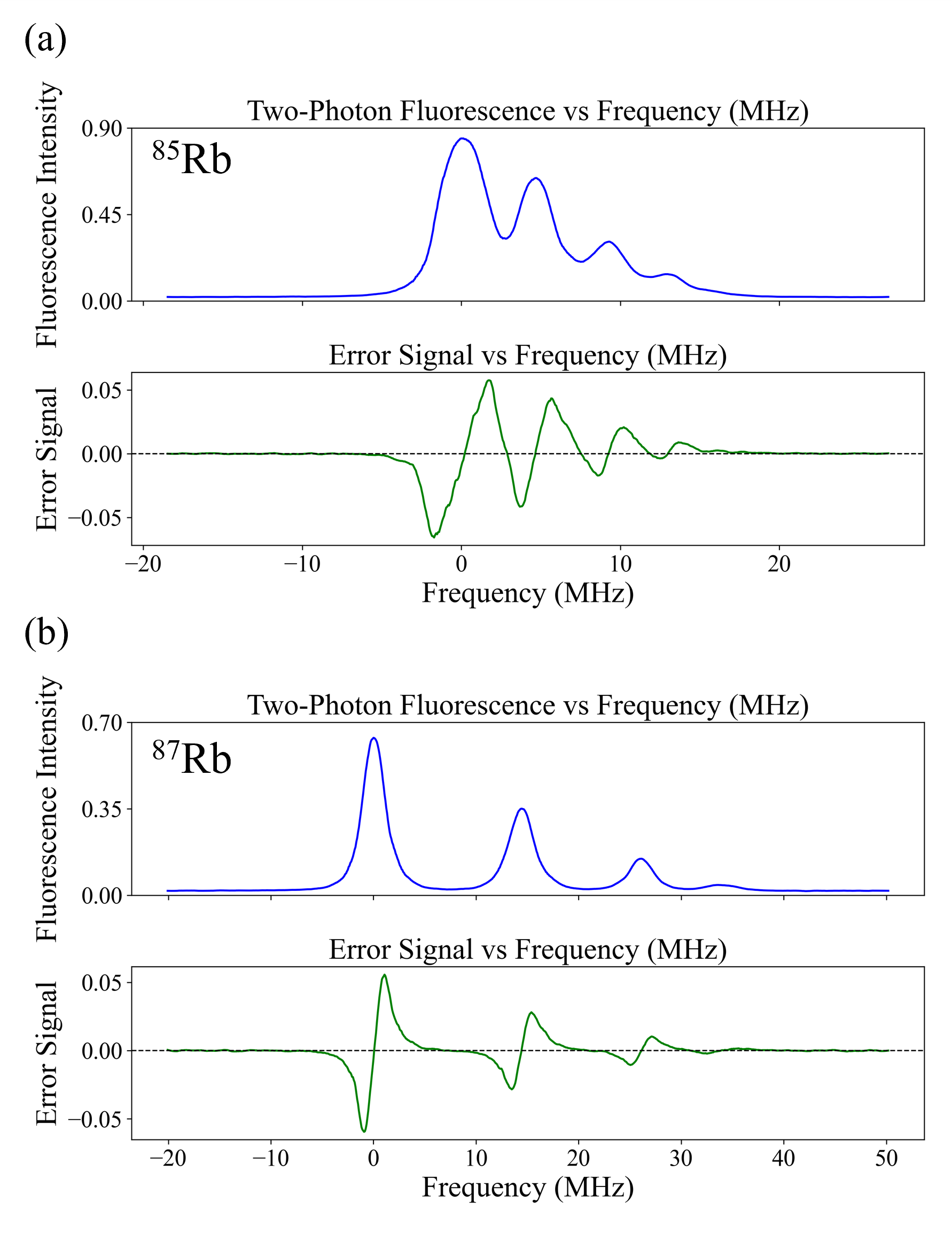}
\caption{(a) (Top) Two-photon Doppler-free spectrum of hot \ce{^{85}_{}Rb} vapor (heated to $\approx 90^\circ C$). FWHM (the strongest hyperfine peak) is 3.8 MHz. (Bottom) Error signal (green) used for locking. The points at which the error signal is 0.00 (dashed line) represent potential lock points. (b) (Top) Two-photon Doppler-free spectrum in hot \ce{^{87}_{}Rb} vapor (heated to $\approx 90^\circ$ C). FWHM (the strongest hyperfine peak) is 2.4 MHz. (Bottom) corresponding error signal.}
\label{fig:Doppler-free-lock}
\end{figure}

The two-photon Doppler-free spectrum and associated error signals used for peak locking are shown in Figure~\ref{fig:Doppler-free-lock}. We achieved two-photon lock linewidths of $<$3.5 kHz (less than the linewidth of the laser grating) for \ce{^{85}_{}Rb} and \ce{^{87}_{}Rb} $5\mathrm{S}_{1/2} \rightarrow 5\mathrm{D}_{5/2}$ two-photon excitation. This lock stability enables probing of our MOT cloud without deviations off resonance.

\subsubsection{Absolute two-photon cross section}
\label{app:details-two-photon-cross section}

For a degenerate CW two-photon transition \(|g\rangle\!\to\!|e\rangle\) at wavelength \(\lambda\),
the detected fluorescence count rate \(F\) (s\(^{-1}\)) under weak excitation is
\begin{equation}
F \;=\; \eta\, B\, \frac{\delta}{(h\nu)^2} \int n(\mathbf{r})\, I^2(\mathbf{r}) \, d^3\mathbf{r},
\label{eq:F_master}
\end{equation}
where \(\eta\) is the total detection efficiency, \(B\) is the branching ratio for the detected channel,
\(\nu=c/\lambda\), \(n(\mathbf{r})\) is the atomic number density, \(I(\mathbf{r})\) is the irradiance,
and \(\delta\) is the (molecular/atomic) two-photon cross section in GM units
(\(1~\mathrm{GM}\equiv 10^{-50}\,\mathrm{cm}^4\,\mathrm{s}\,\mathrm{photon}^{-1}\)).

\paragraph{CW Gaussian beam}
For a fundamental Gaussian beam of total power \(P\) and waist \(w_0\) at the atoms,
\begin{equation}
\begin{aligned}
I(r,z) &= \frac{2P}{\pi w^2(z)}
\exp\!\left(-\frac{2r^2}{w^2(z)}\right), \\[4pt]
w(z) &= w_0\sqrt{1+\left(\frac{z}{z_R}\right)^2}, 
\qquad
z_R = \frac{\pi w_0^2}{\lambda}.
\end{aligned}
\label{eq:gaussian}
\end{equation}
If the cloud is large compared with \(w_0\) transversely and extends over many Rayleigh ranges
(\(|z|\!\gg\! z_R\) where \(I^2\) still contributes), we may take \(n(\mathbf{r})\approx n\) (constant)
over the region where \(I^2\) is non-negligible. Then
\begin{equation}
\int I^2(\mathbf{r})\, d^3\mathbf{r}
= \frac{P^2 z_R}{w_0^2}.
\label{eq:I2_integral}
\end{equation}
More generally, for a finite cloud between \(z=z_1\) and \(z=z_2\),
\begin{equation}
\int I^2(\mathbf{r})\, d^3\mathbf{r}
= \frac{P^2}{\pi}\!
\int_{z_1}^{z_2}\!
\frac{dz}{w^2(z)}
= \frac{P^2 z_R}{\pi w_0^2}
\left[\arctan\!\left(\frac{z}{z_R}\right)\right]_{z_1}^{z_2}.
\label{eq:I2_finite}
\end{equation}
If the cloud is Gaussian with FWHM sizes \(\mathrm{FWHM}_{x,y,z}\), one may approximate the density
by a 3D Gaussian and evaluate \(\int n\,I^2 dV\) as a product of 1D Gaussian integrals near the waist;
for \(\mathrm{FWHM}_z\!\gg\!z_R\) it reduces back to Eq.~\eqref{eq:I2_integral}.

\paragraph{Solving for $\delta$ from a log--log power scan}
Experimentally one fits $F=\alpha\,P_{\mathrm{meas}}^2$ on a log--log plot (slope $\approx 2$ confirms two-photon excitation).
The measured power is related to the power at the atoms by $P_{\mathrm{atoms}}=T\,P_{\mathrm{meas}}$, with $T\!\approx\!1.0254$ (Table~\ref{tab:two-photon-params}) accounting for the beam-splitter ratio and losses. Because $F\propto P^2$, this transmission is absorbed once into the corrected rate prefactor through the input-loss correction $\alpha\to\alpha/T^2$ (Table~\ref{tab:two-photon-params}), so that the fit is expressed in terms of the power at the atoms, $F=\alpha\,P_{\mathrm{atoms}}^2$. Using Eqs.~\eqref{eq:F_master} and \eqref{eq:I2_integral}:
\begin{equation}
\begin{aligned}
\alpha\,P_{\mathrm{atoms}}^2
&= \eta\,B\,
   \frac{\delta}{(h\nu)^2}\,
   n\!\left(
      \frac{P_{\mathrm{atoms}}^2 z_R}{w_0^2}
     \right), \\[6pt]
\delta
&= \alpha\,
   \frac{(h\nu)^2}{\eta\,B\,n}\,
   \frac{w_0^2}{z_R}.
\end{aligned}
\label{eq:delta_from_alpha}
\end{equation}

Here $n$ is the number density. By measuring the total atom number \(N\) and assuming the cloud is near-uniform across
the illuminated region, we substitute \(n \simeq N/V_{\text{cloud}}\). For a 3D Gaussian atom cloud,
\(V_{\text{cloud}}=(2\pi)^{3/2}\sigma_x\sigma_y\sigma_z\), with \(\sigma=\mathrm{FWHM}/(2\sqrt{2\ln2})\).
We also use the beam size conversions: \(w_0=\mathrm{FWHM}_\text{beam}/\sqrt{2\ln2}\).

\paragraph{Minimum statistically significant excitation power and flux}

In a log--log excitation scan the measured fluorescence is converted to
\(y=\log_{10}F\) and plotted versus either \(x=\log_{10}P_{\mathrm{meas}}\) or,
after correcting for the beam geometry, \(x=\log_{10}\Phi\), where \(\Phi\) is
the photon flux at the atoms. The flux is defined as the laser power per unit
beam area (using the $1/e^2$ mode area $\pi w_0^2$) converted to a photon rate,
\begin{equation}
\Phi = \frac{P_{\mathrm{atoms}}}{\pi w_0^2\, h\nu},
\qquad P_{\mathrm{atoms}} = T_{\mathrm{total}}\,P_{\mathrm{meas}},
\label{eq:flux_def}
\end{equation}
where $w_0$ is the $1/e^2$ beam waist radius, $h\nu$ is the photon energy at
$778.1$~nm, and $T_{\mathrm{total}}=1.0254$ (Table~\ref{tab:two-photon-params})
corrects the measured power to the power at the atoms. This power-per-beam-area
convention matches that used by the comparison studies of Table~\ref{tab:tp-flux}.
Only data points whose measured count rates lie
above the experimentally determined noise floor are included in the linear
regression, but the noise characteristics of \emph{all} data points (including
those below the noise floor) are used in determining the detection limit.

Over the region used for fitting, the two-photon response is well described by
the weighted linear model
\begin{equation}
    y(x) = m x + b,
\end{equation}
with parameter covariance
\begin{equation}
    \Sigma =
    \begin{pmatrix}
        \sigma_m^2 & \mathrm{Cov}(m,b)\\
        \mathrm{Cov}(m,b) & \sigma_b^2
    \end{pmatrix}.
\end{equation}
The $1\sigma$ uncertainty of the fitted line follows from standard error
propagation.  With
\[
    \mathbf{g}(x)
    =
    \begin{pmatrix}
        \partial y/\partial m\\[4pt]
        \partial y/\partial b
    \end{pmatrix}
    =
    \begin{pmatrix}
        x\\[4pt]
        1
    \end{pmatrix},
\]
the variance of the fitted mean is
\begin{equation}
    \sigma_{\mathrm{fit}}^2(x)
    =
    \mathbf{g}(x)^{\mathsf{T}}
    \,\Sigma\,\mathbf{g}(x)
    =
    x^2\sigma_m^2 + \sigma_b^2 + 2x\,\mathrm{Cov}(m,b).
\end{equation}

Measurement noise, which grows rapidly as the fluorescence approaches the
noise floor, contributes an additional term \(\sigma_{\mathrm{data}}(x)\).
The total $1\sigma$ predictive uncertainty at excitation level \(x\) is
\begin{equation}
    \sigma_{\mathrm{pred}}(x)
    =
    \sqrt{
        \sigma_{\mathrm{fit}}^2(x)
        +
        \sigma_{\mathrm{data}}^2(x)
    }.
\end{equation}

Let \(y_{\mathrm{noise}}\) denote the measured noise-floor level in log space.
The minimum statistically significant excitation is defined as the smallest
\(x\) for which the lower edge of the prediction band exceeds the noise floor:
\begin{equation}
\begin{aligned}
y\!\left(x_{\mathrm{det}}\right)
-
\sigma_{\mathrm{pred}}\!\left(x_{\mathrm{det}}\right)
&= y_{\mathrm{noise}}.
\end{aligned}
\label{eq:detection_condition}
\end{equation}
Equation~\eqref{eq:detection_condition} is solved numerically using a
bracketing root-finding method.  Because the interpolation
\(\sigma_{\mathrm{data}}(x)\) incorporates all measured points, including those
below the noise floor, no additional constraint is required on the location of
\(x_{\mathrm{det}}\).

Finally, the minimum statistically significant excitation levels in physical
units are
\begin{equation}
    P_{\mathrm{det}}
    = 10^{\,x_{\mathrm{det}}^{(P)}},
    \qquad
    \Phi_{\mathrm{det}}
    = 10^{\,x_{\mathrm{det}}^{(\Phi)}} ,
\end{equation}
which represent the lowest excitation levels at which the two-photon signal is
distinguishable from the noise floor at the \(1\sigma\) confidence level.

\subsubsection{Absorption Imaging}
\label{app:details-absorption-imaging}

\paragraph{Frequency Stabilization and Sweeping}

Absorption imaging was performed using a probe laser (a separate laser) that was frequency stabilized to the trap laser transition for absorption imaging by frequency offset locking (FOL), see reference for a general FOL scheme \cite{Schünemann1999}. In brief, the repump laser was locked to its atomic transition via side-locking. Then, using polarization beam-splitters, the re-pump laser and the probe laser were combined using a 50/50 beam splitter cube (one laser sent through each face). The polarizations of the repump and probe laser were set to be the same in order to ensure proper mixing \cite{Schünemann1999}. The mixed lasers were fiber-coupled and sent for detection (Vescent D2 Beat Note Detector), with a difference in frequency on the order of 2.8–2.9 GHz ($^{85}$Rb) and
6.6 GHz ($^{87}$Rb). The GHz beat note was sent into a high frequency divider chip (ADF4007), and divided down into the MHz regime (by a factor of 64), for comparison to a reference (TPI 1005 synthesizer). The output of the ADF4007 chip was sent into a loop filter. By adjusting the reference frequency (synthesizer reference), the frequency offset was adjusted around the Rb trap transition.

Our absorption imaging scheme was inspired by other literature \cite{Luksch2012}. In brief, our probe laser was spatially filtered and optically chopped out of phase with our MOT trap and repump lasers. The probe laser was then directed through the MOT and imaged on a camera. 

\paragraph{Optical Density Fitting}
\paragraph{Absorption model with fixed natural linewidth}
After normalizing the data, we model the transmitted intensity under the Beer–Lambert law with a Lorentzian line shape of fixed full-width half-maximum (FWHM) $\gamma$ (in $\mathrm{MHz}$). We use $\gamma = 6.0{\bar{6}}~\mathrm{MHz}$ and introduce a small phenomenological asymmetry
parameter \(s\) to account for residual skew in the line:
\begin{align}
\label{eq:model}
I_{\text{model}}(\omega;\,\boldsymbol\theta)
&= a \;+\; \exp\!\left[-\,\mathrm{OD}\;\mathcal{L}(\omega;\,\omega_0,s)\right],\\[4pt]
\mathcal{L}(\omega;\,\omega_0,s)
&= \frac{\gamma^2}{4(\omega-\omega_0)^2 + s(\omega-\omega_0) + \gamma^2}.
\end{align}
Here \(\boldsymbol\theta = (\mathrm{OD},\, \omega_0,\, a,\, s)\), where
\(\mathrm{OD}\) is the on-resonance optical density, \(\omega_0\) is the line
center (in \si{MHz}), \(a\in[0,1]\) is a constant offset, and \(s\) controls weak asymmetry (\(s=0\) recovers the symmetric Lorentzian). We use \(s\) purely as a nuisance parameter to absorb mild
nonidealities (e.g.\ residual gradients or detuning asymmetries).

\paragraph{Nonlinear least squares fit and constraints}

For each pixel \((r,c)\) we minimize the residual sum of squares
\begin{equation}
\label{eq:lsq}
\min_{\boldsymbol\theta}\; \mathrm{RSS}(\boldsymbol\theta)
\;=\; \sum_{i\in\mathcal{K}} \left[
I_{\text{model}}(\omega_i;\,\boldsymbol\theta)
- \hat y_i(r,c)
\right]^2,
\end{equation}
over the index set \(\mathcal{K}\) remaining after filtering and trimming.
We impose simple box constraints reflecting the acquisition and normalization:
\begin{equation}
\begin{aligned}
&0 \le \mathrm{OD} \le 50, \qquad
\min_i \omega_i \le \omega_0 \le \max_i \omega_i,\\
&0 \le a \le 1, \qquad -100 \le s \le 100.
\end{aligned}
\end{equation}

\paragraph{Goodness of fit and acceptance criterion}

We report the coefficient of determination
\begin{equation}
\label{eq:r2}
R^2 \;=\; 1 - \frac{\sum_{i\in\mathcal{K}}\left(\hat y_i-\hat y_i^{\text{fit}}\right)^2}%
{\sum_{i\in\mathcal{K}}\left(\hat y_i-\bar y\right)^2},
\qquad
\bar y \;=\; \frac{1}{|\mathcal{K}|}\sum_{i\in\mathcal{K}} \hat y_i,
\end{equation}
where \(\hat y_i^{\text{fit}} = I_{\text{model}}(\omega_i;\,\hat{\boldsymbol\theta})\).
If the denominator in Eq.~\eqref{eq:r2} vanishes we set \(R^2=-1\).
Fits with \(R^2 < 0.70\) are rejected by assigning \(\widehat{\mathrm{OD}}=0\).

\begin{figure}[!b]
\includegraphics[width=7 cm]{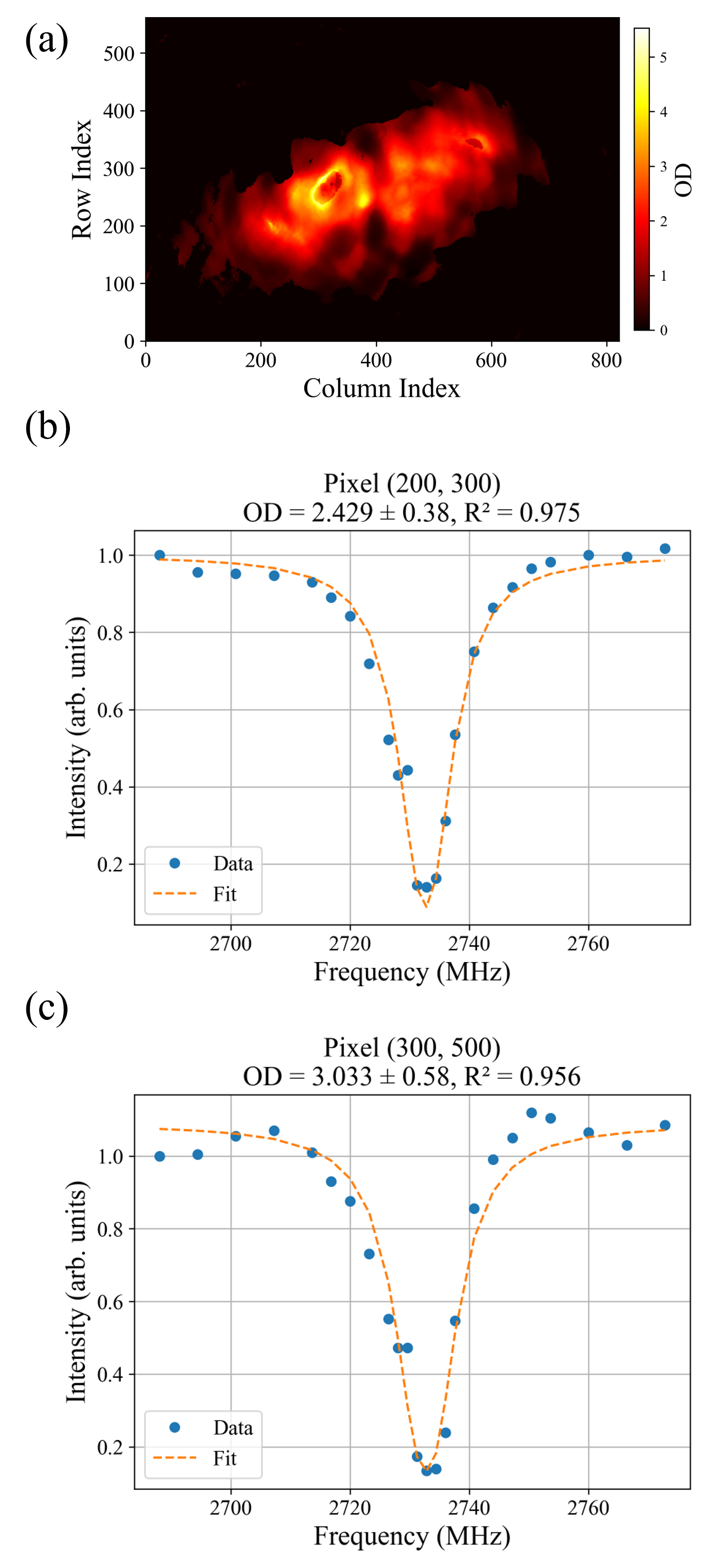}
\caption{(a) MOT absorption image with OD plotted as an intensity surface plot. (b-c) Fitting of example pixels used to produce the absorption OD plot (OD reported at the top of each figure, data shown in blue, fit shown in orange).}
\label{fig:absorption-imaging}
\end{figure}

\paragraph{Optical density map}
We assemble the per-pixel estimate \(\widehat{\mathrm{OD}}(r,c)\) into a 2D map
of size \(H\times W\) matching the input images. For a user-specified single
pixel \((r^\star,c^\star)\), we can additionally plot the data and best-fit
\(\{\omega_i,\hat y_i(r^\star,c^\star)\}\) together with
\(I_{\text{model}}(\omega;\,\hat{\boldsymbol\theta})\) for evaluation (such as in Figure~\hyperref[fig:absorption-imaging]{\ref*{fig:absorption-imaging}(b)} and Figure~\hyperref[fig:absorption-imaging]{\ref*{fig:absorption-imaging}(c)}). All fits use bound-constrained nonlinear least squares (Levenberg--Marquardt-type)
as implemented in \texttt{least\_squares}.

\paragraph{Per-pixel uncertainty}
For each pixel fit, the nonlinear least-squares routine
(\texttt{least\_squares}) returns the Jacobian $J$ of the residuals with
respect to the parameters, evaluated at the best-fit
$\hat{\boldsymbol\theta} = (\widehat{\mathrm{OD}},\hat\omega_0,\hat a,\hat s)$.
Let $r_i(\hat{\boldsymbol\theta})$ denote the residual at frequency
$\omega_i$, and define the residual sum of squares
\begin{equation}
\mathrm{RSS}(\hat{\boldsymbol\theta})
=
\sum_{i\in\mathcal{K}} r_i(\hat{\boldsymbol\theta})^2.
\end{equation}
With $N = |\mathcal{K}|$ data points and $P=4$ fit parameters, we determine the
noise variance as
\begin{equation}
\sigma^2
=
\frac{\mathrm{RSS}(\hat{\boldsymbol\theta})}{N-P}.
\end{equation}
The parameter covariance matrix is then approximated by the usual
Gauss–Newton expression
\begin{equation}
\label{eq:cov}
\Sigma_{\hat{\boldsymbol\theta}}
=
\sigma^2 (J^\mathsf{T} J)^{-1}.
\end{equation}
We take the on-resonance optical-density uncertainty for that pixel to be the
square root of the corresponding diagonal element,
\begin{equation}
\label{eq:od_err}
\delta\mathrm{OD}
=
\sqrt{\Sigma_{\hat{\boldsymbol\theta},11}},
\end{equation}
where index $1$ refers to the $\mathrm{OD}$ parameter in
$\boldsymbol\theta = (\mathrm{OD},\omega_0,a,s)$.
This ``fit-based'' $1\sigma$ uncertainty incorporates photon shot noise,
camera noise, and any residual model mismatch that appears as scatter in the
normalized intensity.
The per-pixel uncertainties $\delta\mathrm{OD}(r,c)$ are stored in a map of
the same size as $\widehat{\mathrm{OD}}(r,c)$ and are used to propagate
errors to the total atom number.
Pixels with $R^2<0.70$ are assigned $\widehat{\mathrm{OD}}=0$ but retain their
formal $\delta\mathrm{OD}$ in the error budget, so they contribute only to the
uncertainty and not to the mean atom number.

\paragraph{Total atom number}
The total atom number $N$ is obtained by summing the optical density over all
pixels and converting to atom number using the resonant scattering cross
section,
\begin{equation}
\label{eq:N_from_OD}
N
=
\frac{A_{\text{pixel}}}{\sigma_0}
\sum_{r,c} \widehat{\mathrm{OD}}(r,c),
\qquad
\sigma_0 = \frac{3\lambda^2}{2\pi},
\end{equation}
where $A_{\text{pixel}}$ is the effective pixel area in the object plane
(camera pixel area divided by the square of the imaging magnification) and
$\lambda$ is the imaging wavelength.  In the analysis code we implement
Eq.~\eqref{eq:N_from_OD} using a scale factor
\(\text{scale} = A_{\text{pixel}}/\sigma_0\).

Assuming that the per-pixel OD estimates are statistically independent, the
variance of the summed OD is
\begin{equation}
\label{eq:var_OD_sum}
\mathrm{Var}\!\left[\sum_{r,c} \widehat{\mathrm{OD}}(r,c)\right]
=
\sum_{r,c} \delta\mathrm{OD}(r,c)^2,
\end{equation}
so that
\begin{equation}
\mathrm{Var}[N]
=
\left(\frac{A_{\text{pixel}}}{\sigma_0}\right)^2
\sum_{r,c} \delta\mathrm{OD}(r,c)^2.
\end{equation}
We therefore quote
\begin{equation}
\label{eq:N_err}
N
=
N_{\text{est}} \pm \delta N,
\qquad
\delta N
=
\frac{A_{\text{pixel}}}{\sigma_0}
\sqrt{\sum_{r,c} \delta\mathrm{OD}(r,c)^2},
\end{equation}
as the $1\sigma$ statistical uncertainty on the atom number inferred from the absorption image.

\newpage
\section{TWO-PHOTON DETUNINGS}
\label{app:two-photon-detunings}

The following table reports two-photon detunings for various alkali atoms (Li, Na, K, Rb, and Cs).

\begin{table}[h!]
\centering
\caption{Degenerate two-photon wavelengths and detunings of the virtual state from the nearest real state for Li, Na, K, Rb, and Cs. A positive detuning indicates that the virtual state lies above the nearest real state (i.e., the drive laser is blue detuned from that intermediate state); a negative detuning indicates red detuning. The Rb two-photon transition studied in this manuscript is in bold.}
\renewcommand{\arraystretch}{2}
\begin{tabularx}{\linewidth}{X X X X}
    \hline\hline
    \textbf{Atom} & \textbf{Transition} & \textbf{$\lambda$ (nm)} & \textbf{Detuning (THz)}\\
    \hline
    Li \cite{DeGraffenreid2003PRA, DeGraffenreid2003JPhysB} & 2S $\rightarrow$ 4S & 571.3 & +78.0\\
    Li \cite{Tisone1987, Labazan2000} & 2S $\rightarrow$ 3D & 639.1 & +22.1\\
    Na \cite{Levenson1974} & 3S $\rightarrow$ 5S & 602.2 & -10.7\\
    Na \cite{Haensch1974} & 3S $\rightarrow$ 4D & 578.7 & +9.0\\
    K \cite{Liu2001} & 4S $\rightarrow$ 6S & 728.4 & +20.5\\
    K \cite{Liu2001} & 4S $\rightarrow$ 4D & 730.0 & +19.7\\
    Rb \cite{Ko2004,Morzynski2013,Wang2019} & 5S $\rightarrow$ 7S & 760.0 & +10.2\\
    \textbf{Rb} \cite{Saha2011,Zapka1983} & \textbf{5S $\rightarrow$ 5D} & \textbf{778.1} & \textbf{+1.06}\\
    Cs \cite{Nunez2023} & 6S $\rightarrow$ 8S & 822.2 & +13.0\\
    Cs \cite{Georgiades1994} & 6S $\rightarrow$ 6D & 883.7 & +4.13\\
    \hline\hline
\end{tabularx}
\label{tab:alkali-two-photon}
\end{table}

\clearpage

\section{TWO-PHOTON EXCITATION LINEWIDTHS AND OPTICAL CHOPPING}
\label{app:MOT-excitation-spectra}

\begin{figure}[!h]
\includegraphics[width=\linewidth]{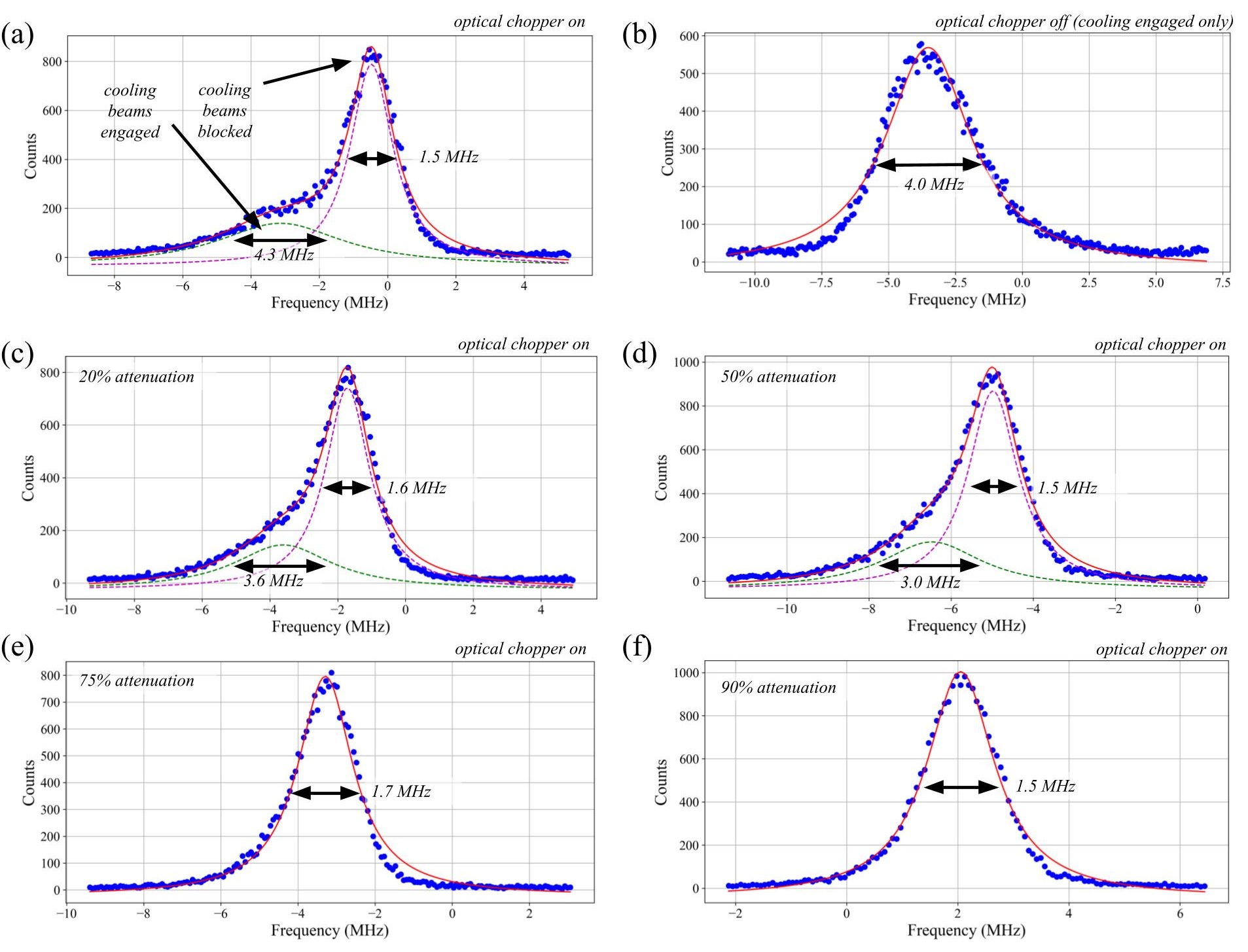}
\caption{MOT excitation spectra for various conditions, (a) Excitation spectra under full trap and repump power (optical chopper engaged), (b) Excitation spectra under full trap and repump power (optical chopper disengaged, therefore only the power broadening component exists), (c-f) Excitation spectra under attenuated trap and repump powers (optical chopper engaged).}
\label{fig:MOT-excitation-spectra}
\end{figure}

\indent To excite the ultracold Rb MOT via two-photon excitation, the trap and repump beams must be blocked. If an optical chopper is not used, a significant portion of the atom population will be in the 5P$_{3/2}$ state. From a theoretical standpoint, this is not ideal for modeling of the two-photon transition (although basic models incorporating the occupied 5P$_{3/2}$ state have been considered for metrological applications) \cite{Leng2019}.  In addition, with the trap and repump beams engaged, line-broadening (by a factor of $>$250\%) is observed, as shown in Figure~\hyperref[fig:MOT-excitation-spectra]{\ref*{fig:MOT-excitation-spectra}(a)} with the $^{87}$Rb isotope studied. In the case of using full-power trap and repump beams, in the absence of optical chopping, a 4.0 MHz linewidth is observed (Figure~\hyperref[fig:MOT-excitation-spectra]{\ref*{fig:MOT-excitation-spectra}(b)}).

As discussed for the two-photon transitions of Li \cite{Sautenkov2017, Sautenkov2018} and Cs \cite{Georgiades1994} atoms cooled via MOT, broadening when the cooling beams are engaged is due to power-broadening from the cooling lasers. This is proven by attenuating the cooling laser power (Figure~\hyperref[fig:MOT-excitation-spectra]{\ref*{fig:MOT-excitation-spectra}(f)}). At 90\% attenuation a single Lorentzian that matches the linewidth with the cooling beams blocked is observed. For the main experiments in this paper (i.e. Figure~\ref{fig:log-log}), the two-photon excitation is time-gated and the data are analyzed only when the cooling beams are blocked.

\onecolumngrid

\twocolumngrid
\onecolumngrid
\clearpage
\section{FULL LIST OF TWO-PHOTON PARAMETERS TO CALCULATE THE CROSS SECTION}

See Table~\ref{tab:two-photon-params}.

\label{app:two-photon-params}
\begin{table}[h]
\centering
\caption{Quantities required to obtain the TPA cross sections for $^{85}$Rb and $^{87}$Rb.
Excitation-power correction uses the measured chain: beam splitter mismatch, pre-cell optics, and a single glass face.
Uncertainty budgets for $\delta$ combine (in quadrature) detection efficiency, atom number, fit-intercept, and beam-size contributions:
$^{85}$Rb (12.4\%, 8.5\%, 6.2\%, 10.0\%) $\Rightarrow$ 19.1\%;
$^{87}$Rb (12.4\%, 9.5\%, 1.2\%, 10.0\%) $\Rightarrow$ 18.6\%. The power y-intercept values below were used to calculate the TPA cross section, though the flux y-intercept values reported in Fig.~\ref{fig:log-log} could also be used (for Fig.~\ref{fig:log-log} the input-loss correction is applied since it is a report of the flux values at the atoms). We note that the flux parameter is a combination of the two measured parameters of the laser power and spot size. Taking this into account, the flux y-intercept can be derived from the power y-intercept and the calculated two-photon cross section would be consistent with either approach.}
\setlength{\tabcolsep}{6pt}
\renewcommand{\arraystretch}{1.12}
\begin{tabular}{@{} l c l @{}}
\hline\hline
\textbf{Parameter} & \textbf{Symbol} & \textbf{Value} \\
\hline
\multicolumn{3}{@{}l}{\textit{Beam / Mode (common to both isotopes)}} \\
Beam FWHM at focus & $\mathrm{FWHM}$ & $7.64 \pm 0.19~\mu\mathrm{m}$ \\
Beam waist radius & $w_0$ & $6.49 \pm 0.16~\mu\mathrm{m}\;(=6.489\times10^{-4}~\mathrm{cm})$ \\
Rayleigh length & $z_R$ & $0.170 \pm 0.009~\mathrm{mm}$ \\
\addlinespace[3pt]
\multicolumn{3}{@{}l}{\textit{Detection / Throughput (common)}} \\
Overall detection efficiency & $\eta$ & $0.0426 \pm 0.0053$ \\
\quad PMT quantum efficiency & $\eta_{\mathrm{PMT}}$ & $0.284$ \\
\quad Collection optics efficiency & $\eta_{\mathrm{optics}}$ & $0.161 \pm 0.020$ \\
\quad Fluorescence filters (96\% \& 97\%) & $\eta_{\mathrm{filters}}$ & $0.96\times0.97=0.9312$ \\
Branching ratio to detected line & $B$ & $0.08$ \;(5D$\!\to\!$6P$_{3/2}\!\sim\!30\%$; 6P$_{3/2}\!\to\!$5S$_{1/2}$ 420 nm $\sim\!25\%$) \cite{Sheng2008} \\
\addlinespace[3pt]
\multicolumn{3}{@{}l}{\textit{Excitation Power Path (meter $\to$ atoms, common)}} \\
beam splitter mismatch & $S_\mathrm{split}$ & $210.6/191.0 = 1.1026$ \\
Pre-cell optics & $T_\mathrm{opt}$ & $201.5/210.6 = 0.9568$ \\
Single glass face (circular pol.) & $T_\mathrm{win}$ & $0.972$ \\
Net transmission & $T_\mathrm{total}$ & $S_\mathrm{split}T_\mathrm{opt}T_\mathrm{win}=1.0254$ \\
Input-loss correction (excitation) & \textnormal{corr} & $\alpha\to\alpha/1.0515$ \\
\hline
\multicolumn{3}{@{}l}{\textbf{$^{85}$Rb (TPA measurement)}} \\
Total atom number & $N$ & $(1.37 \pm 0.12)\times10^{7}$ \\
Cloud FWHM $(x,y,z)$ & $\mathrm{FWHM}_{x,y,z}$ & $(1.768,\;0.462,\;0.626)~\mathrm{mm}$ \\
Peak density & $n_0$ & $(2.22 \pm 0.19)\times10^{10}~\mathrm{cm}^{-3}$ \\
Free fit (base-10 logs) & $m,\,b$ & $m=2.041 \pm 0.023,\;\;b=-0.279 \pm 0.027$ \\
Fitted rate prefactor & $\alpha$ & $(5.00 \pm 0.31)\times10^{11}~\mathrm{s}^{-1}\,\mathrm{W}^{-2}$ \\
Two-photon cross section & $\delta$ & $(1.07 \pm 0.20)\times10^{-38}~\mathrm{cm}^4\!\cdot\!\mathrm{s}/\mathrm{photon}$ \\
& & $(1.07 \pm 0.20)\times10^{12}~\mathrm{GM}$ \\
\hline
\multicolumn{3}{@{}l}{\textbf{$^{87}$Rb (TPA measurement)}} \\
Total atom number & $N$ & $(7.02 \pm 0.67)\times10^{6}$ \\
Cloud FWHM $(x,y,z)$ & $\mathrm{FWHM}_{x,y,z}$ & $(1.278,\;0.694,\;0.694)~\mathrm{mm}$ \\
Peak density & $n_0$ & $(9.46 \pm 0.90)\times10^{9}~\mathrm{cm}^{-3}$ \\
Free fit (base-10 logs) & $m,\,b$ & $m=2.018 \pm 0.004,\;\;b=-0.256 \pm 0.005$ \\
Fitted rate prefactor & $\alpha$ & $(5.27 \pm 0.06)\times10^{11}~\mathrm{s}^{-1}\,\mathrm{W}^{-2}$ \\
Two-photon cross section & $\delta$ & $(2.64 \pm 0.49)\times10^{-38}~\mathrm{cm}^4\!\cdot\!\mathrm{s}/\mathrm{photon}$ \\
& & $(2.64 \pm 0.49)\times10^{12}~\mathrm{GM}$ \\
\hline\hline
\end{tabular}
\label{tab:two-photon-params}
\end{table}

\clearpage
\twocolumngrid
\FloatBarrier

\section{ALLAN DEVIATION BENCHMARKS}
\label{app:allan-deviation}

Allan deviation measurements, originally conceived during the development of the optical clock \cite{Allan1966}, are a way of benchmarking experimental stability when taking long-term measurements. In past work \cite{Parzuchowski2021}, we have used the Allan deviation to benchmark our experiments. Since in this paper we collect data over extended periods, we performed Allan deviation analysis to ensure statistical robustness. 

\begin{figure}[h!]
\includegraphics[width=7 cm]{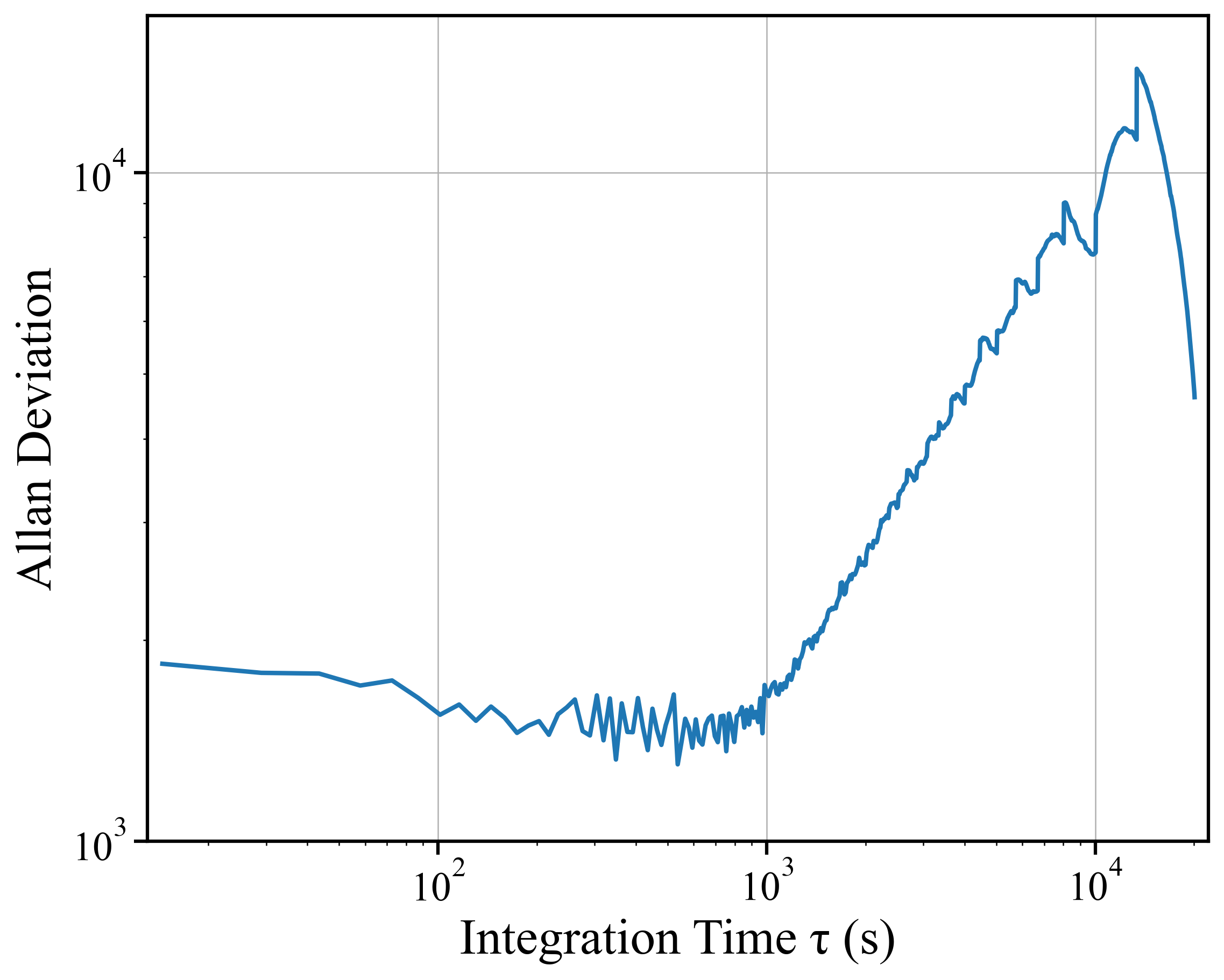}
\caption{Allan deviation of the PMT counts under two-photon excitation (far above the noise floor) of the MOT. The Allan deviation hits a minimum at $\sim$400 s, which sets the block averaging value in the study.}
\label{fig:allan-mot}
\end{figure}

\begin{figure}[h!]
\includegraphics[width=7 cm]{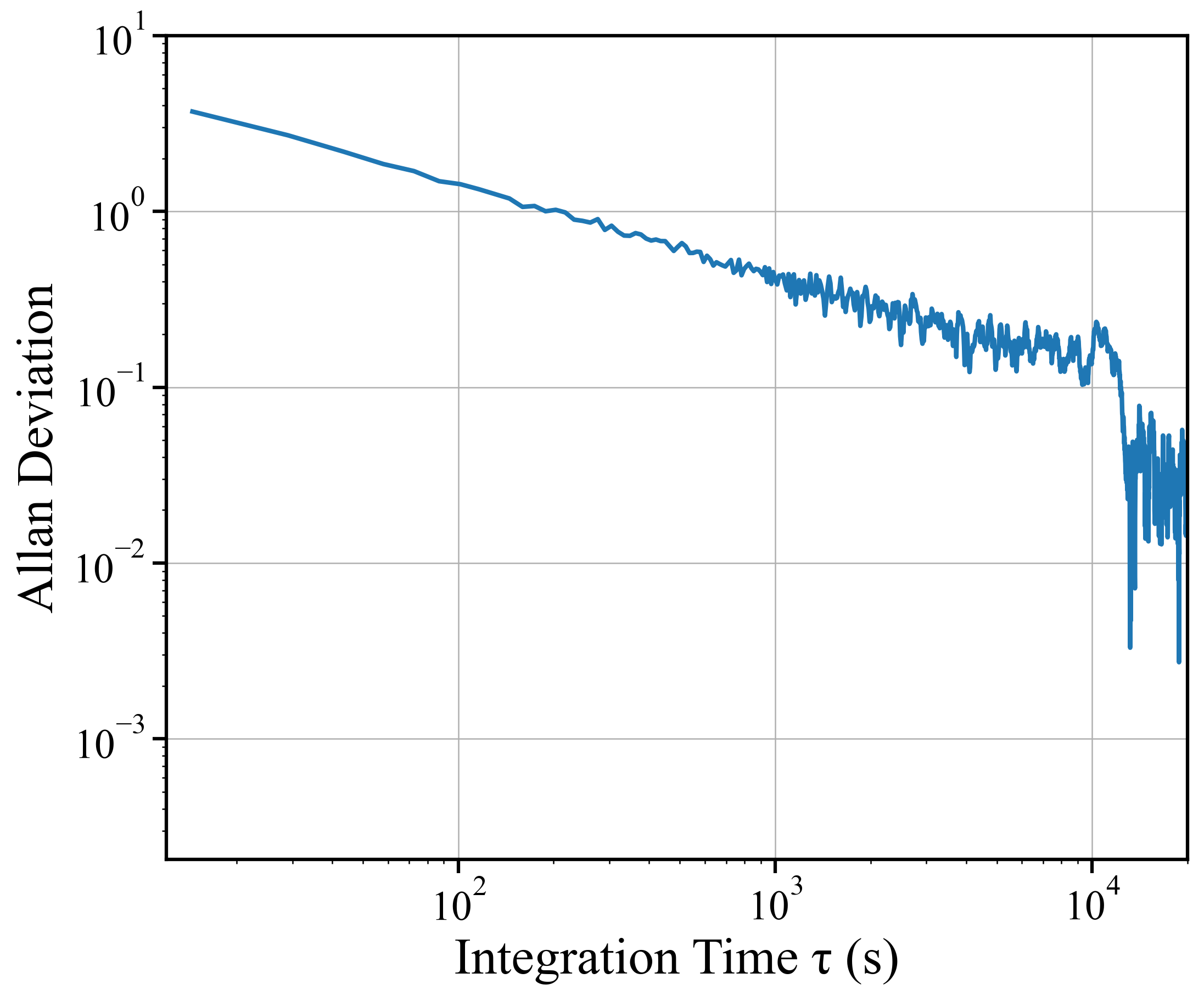}
\caption{Allan deviation of the dark PMT counts with trap and repump beams engaged (the time-gated two-photon component with the trap and repump beams blocked was analyzed). The Allan deviation continues to decrease linearly until the 1 - 1.5 x $10^{4}$ second range, setting the maximum collection time used.}
\label{fig:allan-dark}
\end{figure}

\FloatBarrier
\newpage

\section{TWO-PHOTON LOG-LOG PLOTS OF POWER DEPENDENCE}
\label{app:log-log-power}

\begin{figure}[h!]
\includegraphics[width=7.5cm]{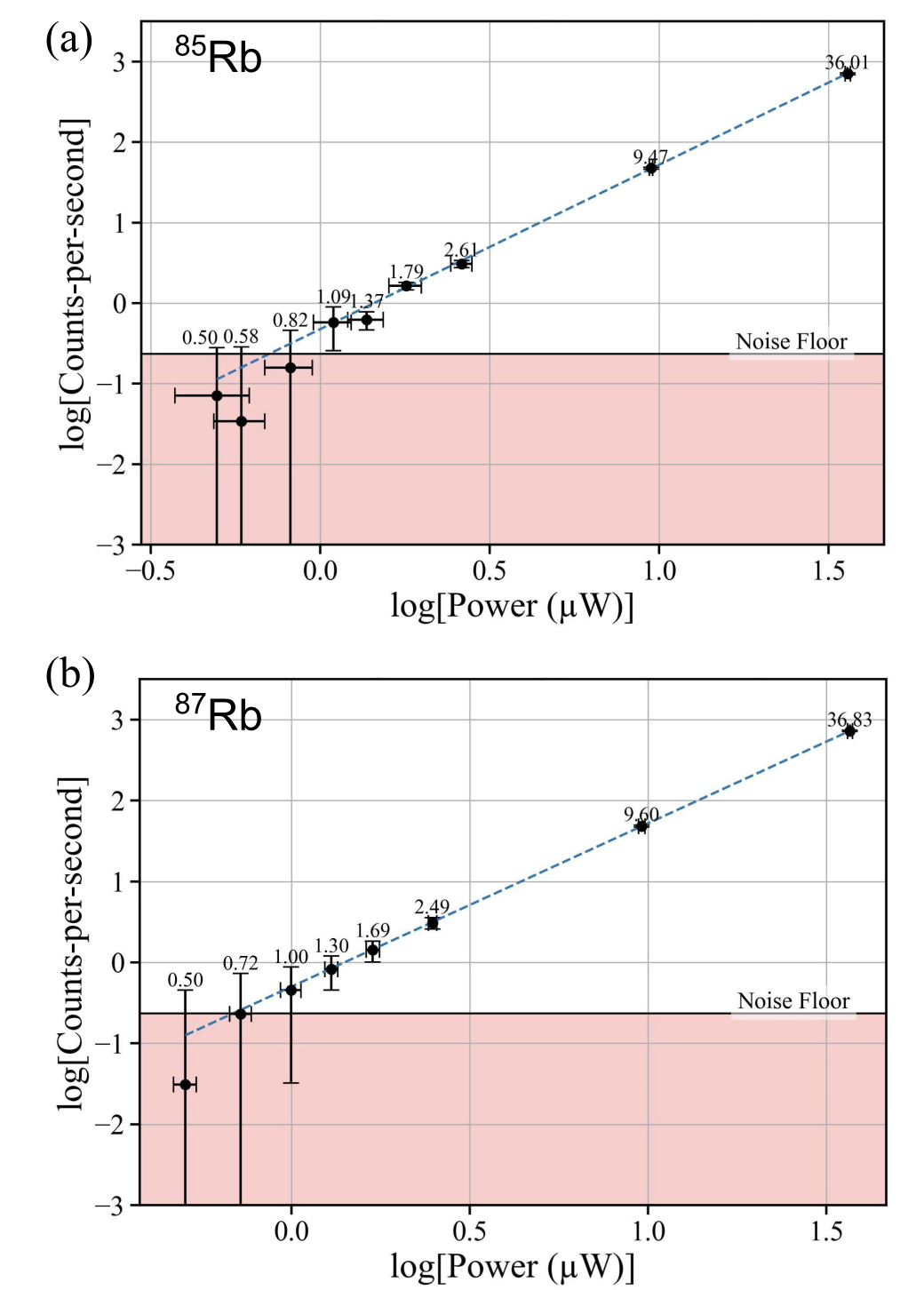}
\caption{Two-photon log-log plot with counts as a function of power. (a) Two-photon power dependence of $^{85}$Rb. (b) Two-photon power dependence of $^{87}$Rb. }
\label{fig:power-dependence}
\end{figure}

Figure~\ref{fig:power-dependence} shows the two-photon log-log plots in terms of laser power (instead of flux). The laser power in $\mu$W is reported above each datapoint for ease of visualization. Using statistical analysis, the lowest significant powers were $1.01_{-0.32}^{+0.46}$ $\mu$W for $^{85}$Rb and $1.12_{-0.45}^{+0.74}$ $\mu$W for $^{87}$Rb.

\FloatBarrier
\clearpage

\section{FLUORESCENCE COLLECTION EFFICIENCY}
\label{app:fluorescence-efficiency}

\begin{figure}[!h]
\includegraphics[width=\linewidth]{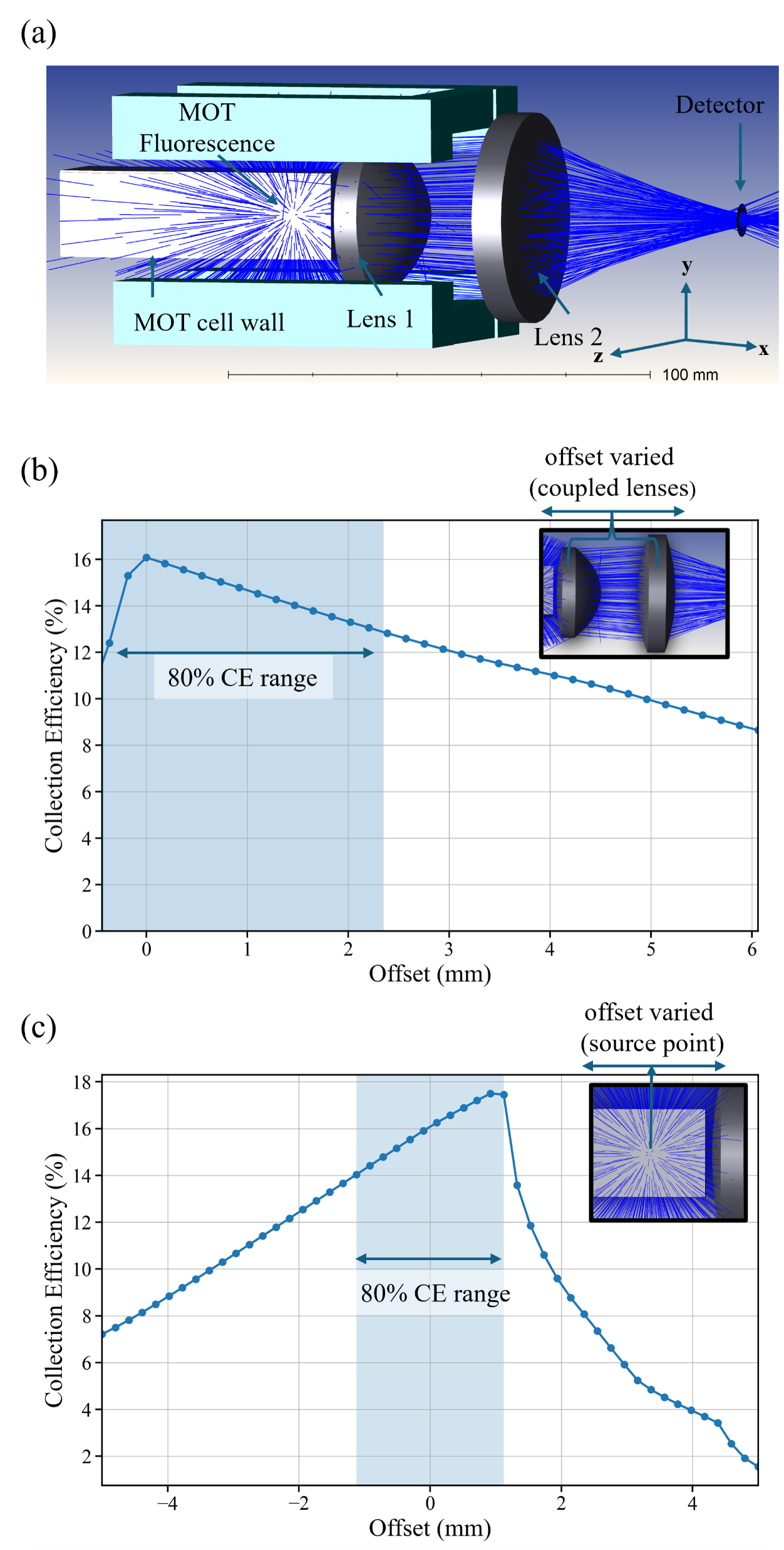}
\caption{(a) Overview of the lens system (source point, lenses, detector, and MOT apparatus) used to collect TPEF from Rb, (b) Collection efficiency (\%) versus the offset of the 2-lens doublet (the distance between the two lenses in the doublet is fixed to the experimental value). With optimal alignment, 16.1\% fluorescence collection efficiency is achieved. The region of 80\% collection efficiency from the maximum is shown in blue. (c) Collection efficiency (\%) versus the source point offset (0 mm is the measured MOT position). The region of 80\% collection efficiency from the maximum is shown in blue. For (b) and (c) figures the detector position is fixed at the experimental value.}
\label{fig:collection-efficiency}
\end{figure}

A determination of the fluorescence collection efficiency (CE) of our system was necessary (for example, CE is required to determine the TPA cross section). However, characterizing the MOT fluorescence collection through use of the MOT’s own fluorescence has challenges. A one-photon fluorescence signal is not possible to deconvolute from MOT beam scatter and excitation beam scatter. Therefore, we determined the collection efficiency computationally using Zemax. Zemax is a software package that allows the spatial modeling of an optics system in a CAD-like environment, utilizing source, detection, optical, and passive objects \cite{Zemax2022}. Light propagation is modeled using ray tracing or wave propagation methods, with the ray tracing option providing fast results for macroscopic optical systems \cite{Zemax2022}. Components such as the cell walls, coil arms and coil housings are included in our model, Figure~\hyperref[fig:collection-efficiency]{\ref*{fig:collection-efficiency}(a)}. Given physical constraints due to the Rb cell and coils of the MiniMOT V2, Zemax was useful in selecting compatible optics and optimizing their positioning. Before settling on final components, manufacturer-supplied Zemax models were explored for various commercial optics. At the end, we selected a 38.4 mm diameter collimating lens because this was the largest lens that fit in the enclosure to maximize ray solid angle.

To perform the characterization, optics elements were downloaded and imported into the Zemax software in non-sequential mode. The fluorescence source was modeled using a source point object, 10$^{3}$ rays were used for the display (Figure~\hyperref[fig:collection-efficiency]{\ref*{fig:collection-efficiency}(a)}) and 10$^{6}$ rays were used for the computation. The detector was simulated by creating a 2D detector surface object with a minimum aperture 4.0 mm radius (matching the experimental detector size of 8 mm on the Hamamatsu PMT). Maximum collection efficiency was 16.1\% (determined by dividing the total hits on the detector by the input number of rays, 10$^{6}$), Figure~\hyperref[fig:collection-efficiency]{\ref*{fig:collection-efficiency}(b)}. Lens 1 and lens 2 are experimentally coupled translationally. Varying their offset in Zemax showed that collection efficiency is sensitive on the mm scale, but micron-level offsets (controllable using micrometers) allow the maximum collection efficiency to easily be attained as shown in Figure~\hyperref[fig:collection-efficiency]{\ref*{fig:collection-efficiency}(b)}. Similarly, in Figure~\hyperref[fig:collection-efficiency]{\ref*{fig:collection-efficiency}(c)}, the location of the fluorescent point source was varied (with the experimentally measured value set to 0). Though spatial sensitivity does exist, offsets up to the range of approximately 2 mm retain collection efficiency within 80\% of the maximum (of far greater magnitude than the experimentally controllable offset).

\FloatBarrier
\clearpage

\section{LINEWIDTH BROADENING} 
\label{app:broadening-calcs}

This appendix contains linewidth broadening calculations for $^{85}$Rb and $^{87}$Rb.

To calculate the Doppler broadening, we assume that the MOT has obtained the Doppler limit of \SI{145.57}{\micro K}, which is identical for $^{85}$Rb and $^{87}$Rb \cite{SteckRb85, SteckRb87}. We then find the Doppler broadening:
\begin{equation}
    \Delta \nu_D = 2\nu_0 \sqrt{\frac{2k_BT\ln(2)}{mc^2}}
\end{equation}

To find the magnetic broadening, we consider the ``anomalous" Zeeman splitting energies
\begin{equation}
    \Delta E = g_J\mu_BB
\end{equation}
where $\mu_B$ is the Bohr magneton, $g_J$ is the Land\'e $g$ factor, and $B$ is the strength of the magnetic field. To find the maximum broadening, we consider the maximum field strength given the gradient of \SI{14}{G / \cm} (based on the miniMOT specs for the 1A current that is run) and assume excitation only happens within the Rayleigh length of $2z_R$. This is denoted as the Zeeman broadening in Table~\ref{tab:linewidth-broadening}. We use this same equation to find the broadening from the finite local field in the MOT, which our calculation determines to be \SI{0.13}{G}. This value is reasonable in the absence of using correction gradients to shim the MOT center. This broadening is treated separately and is denoted as local field broadening in Table~\ref{tab:linewidth-broadening}.

Pressure broadening was calculated using the standard formula of
\begin{equation}
    \Delta \nu_P = bp
\end{equation}
where $b$ is an experimentally determined coefficient. For $^{85}$Rb and $^{87}$Rb, it has a value of \SI{10 \pm 1}{MHz / Torr} \cite{Sargsyan21}.

Power broadening was estimated using the time-energy uncertainty equation
\begin{equation}
    \Delta E \Delta t \ge \hbar
\end{equation}
from which the estimate of
\begin{equation}
    \Delta \nu \sim \frac{\mu E}{2\pi h}
\end{equation}
can be derived, where $E$ is the electric field strength and $\mu$ is the dipole transition moment. These calculations used reduced matrix elements \cite{eicher1975} to provide an approximate upper bound on the magnitude of the broadening for the highest reported power of \SI{37}{\micro \watt}.

The results of these calculations are found in Table~\ref{tab:linewidth-broadening}. The total broadening includes the natural lifetime broadening of the transition and makes the standard assumption that broadenings add in quadrature \cite{Truong16}.

\begin{table}[h!]
    \renewcommand{\arraystretch}{1.35}
    \caption{Linewidth broadening calculations for Zeeman, Doppler, Local field, and Power broadening mechanisms taking into account parameters from our experiment.}
    \label{tab:linewidth-broadening}
    \begin{tabular}{c c c}
        \hline\hline
        \textbf{Broadening Type} & \textbf{$^{85}$Rb} & \textbf{$^{87}$Rb}  \\
        \hline
        Lifetime Broadening (kHz) & 660 & 660\\
        Zeeman Broadening (kHz) & 350 & 350 \\
        Local field Broadening (kHz) & 1300 & 1300\\
        Doppler Broadening (kHz) & 360 & 370 \\
        Pressure Broadening (Hz) & $< 1$ & $< 1$\\
        Power Broadening (kHz) & $450$ & $450$ \\ 
        \hline
        Total Broadening (MHz) & 1.61 & 1.61\\
        \hline\hline
    \end{tabular}
\end{table}

\clearpage
\bibliography{bib}
\end{document}